\documentclass[journal]{IEEEtran}
\usepackage{cite}
\usepackage{multicol,lipsum}
\usepackage{amsmath,amssymb,amsfonts}
\usepackage{algorithmic}
\usepackage{graphicx}
\usepackage{textcomp}
\usepackage{xcolor}
\usepackage{booktabs}
 \usepackage{hyperref}
\usepackage{enumerate}
\usepackage{algorithm}
\usepackage{multirow}
\usepackage{standalone}
\usepackage{tikz}
\usetikzlibrary{positioning}
\usetikzlibrary{decorations.markings}
\usepackage{caption}
\usepackage{subcaption}
 \usepackage{makecell}
\usepackage{mathrsfs}

\ifCLASSINFOpdf
\else
\fi
\begin{document}
%
\title{ Joint Sparse  Graph   for Enhanced MIMO-AFDM Receiver Design }
%
%
%

\author{Qu Luo, ~\IEEEmembership{Member,~IEEE,}
 Jing Zhu, ~\IEEEmembership{Member,~IEEE,}
 Zilong Liu, ~\IEEEmembership{Senior Member,~IEEE,}
 Yanqun Tang, 
Pei Xiao, ~\IEEEmembership{Senior Member,~IEEE,}
 Gaojie Chen, ~\IEEEmembership{Senior Member,~IEEE,}
and
 Jia Shi, ~\IEEEmembership{Member,~IEEE.}

   \thanks{
 
 Qu  Luo, Jing Zhu, Pei Xiao and Gaojie Chen   are  with the  5G \& 6G  Innovation Centre, University of Surrey, U. K. (email: \{q.u.luo, j.zhu, p.xiao, gaojie.chen \}@surrey.ac.uk).  
Zilong   Liu  is   with   the   School   of   Computer   Science   and   Electronics   Engineering,   University   of   Essex,   U. K. (email:  \ zilong.liu@essex.ac.uk).
Yanqun Tang is with the School of Electronics and Communication Engineering, Sun Yat-sen University, China (email: tangyq8@mail.sysu.edu.cn).
Jia Shi is with the School of Telecommunications Engineering, Xidian University, Xi’an, China (email: jiashi@xidian.edu.cn).  }

 }

%



\maketitle

\begin{abstract}
Affine frequency division multiplexing (AFDM) is a promising chirp-assisted multicarrier waveform for future    high-mobility communications. This paper is devoted to enhanced receiver design for multiple-input–multiple-output AFDM (MIMO-AFDM) systems. Firstly, we introduce a unified variational inference (VI) approach  to approximate the target posterior distribution, under which  the belief propagation (BP) and expectation propagation (EP)-based algorithms are derived. As both VI-based detection and low-density parity-check (LDPC) decoding can be expressed by   bipartite graphs in MIMO-AFDM systems, we construct a joint sparse graph (JSG) by merging the graphs of these two for low-complexity receiver design.  Then, based on this graph model, we present the detailed message propagation of the proposed JSG.   Additionally, we propose an enhanced JSG (E-JSG) receiver based on the linear constellation encoding model. The proposed E-JSG eliminates the need for interleavers, de-interleavers, and log-likelihood ratio transformations, {thus leading to concurrent    detection and decoding  } over the integrated sparse graph.   To  further reduce detection complexity, we introduce a sparse channel method by approaximating  multiple graph edges with insignificant channel coefficients into a single edge on the VI graph. Simulation results show the superiority of the proposed receivers in terms of computational complexity, detection and decoding latency, and error rate performance compared to the conventional ones.

\end{abstract}


%
\IEEEpeerreviewmaketitle

         \vspace{-0.5em}

\section{Introduction}
         \vspace{-0.5em}

 \subsection{Background}
\IEEEPARstart{N}{ext} generation wireless systems and standards e.g., beyond 5G, 6G  are envisioned to provide a wide range of data services, including  ultra-reliable, high-rate, and low-latency communications in   { highly dynamic environments, such as  vehicle-to-everything, unmanned aerial vehicles, high-speed trains, and low-earth-orbit satellite systems, etc. \cite{Toward,9779322}. In these scenarios, one often needs to deal with the rapidly time-varying      channels }\cite{otfsHIGH,zhou2023overview}.
The legacy  multicarrier systems primarily rely on orthogonal frequency division multiplexing (OFDM), thanks to its advantages (e.g., efficient hardware implementation, capability of intersymbol interference mitigation) in linear time-invariant channels \cite{BroadbandOFDM}. Nevertheless, OFDM is sensitive to carrier frequency offsets and Doppler shifts/spreads in rapidly time-varying channels and hence may suffer from drastically degraded error performances \cite{otfsHIGH}. 

Extensive efforts have been dedicated  to designing novel modulation waveforms to support reliable communications in high mobility channels.   For example,  orthogonal time–frequency space (OTFS) modulation \cite{otfsHIGH,RavitejaOTFS} and orthogonal chirp division multiplexing (OCDM)  \cite{OrthogonalcdmOuyang,OCDMPerformance} have attracted   substantial research attention.  The former modulates the information symbols in the delay-Doppler  domain instead of the conventional time–frequency  domain. By doing so, one can  effectively convert  a time-variant channel into a two-dimensional (2D) quasi-static channel in delay-Doppler domain \cite{otfsHIGH,YuanOTFS1,RavitejaOTFS}. 
In contrast, OCDM, building  upon the discrete Fresnel transform (DFnT), modulates the information symbols using a class of orthogonal chirp subcarriers (SCs) \cite{OrthogonalcdmOuyang}.
Both  OTFS and OCDM have demonstrated superior     error performances   over OFDM  \cite{Haifocdm,AFDMTWCBemani}. 
However, the 2D structure of OTFS requires radical change of the transceivers (compared to that of OFDM), whilst it suffers from large overhead for sending pilots and multiuser signals \cite{AFDMTWCBemani}. OCDM, on the other hand, may not be able to achieve optimal   performance when its chirp rate does not fit the specific   delay-Doppler profile of the channel \cite{AFDMTWCBemani}.

 {Recently, Affine }frequency division multiplexing (AFDM) is another    appealing solution for efficient and reliable communications over high-mobility channels \cite{AFDMTWCBemani}. Similar to   OCDM, information symbols in AFDM are   multiplexed on a number  of orthogonal chirp SCs through  the  discrete
affine Fourier transform (DAFT) \cite{ErsegheAFDM}.  By appropriately adjusting the chirp rate based on the delay-Doppler profile of the channel, AFDM enables a separable and quasi-static channel representation, thereby achieving full diversity over doubly selective channels \cite{AFDMTWCBemani}. Since the  DAFT  employed in AFDM is a generalization of many other transforms, such as the discrete Fourier transform (DFT)  and DFnT, AFDM generalizes  OFDM and OCDM as special cases.  
More importantly, AFDM enjoys a high compatibility to the legacy OFDM as DAFT can be efficiently implemented through fast Fourier transform (FFT) with additional two one-tap filters   \cite{BemaniAFDM_conf1}. These advantages make AFDM as  {a promising candidate  multicarrier waveform  for the next generation 
   wireless systems }\cite{NiAFDMSens,savaux2023dft}.

         \vspace{-1em}
 \subsection{Related Works} 

 AFDM has attracted a plethora of research attempts from various aspects, such as system impairments \cite{BemaniAFDM_conf1,tang2023time}, channel estimation \cite{YinChannel,yin2022channel},  signal detection \cite{bemani2022low,wu2023message},  and  intergrated sensing and communications \cite{zhu2023low,ni2022afdm,IntegratedAFDM}.
The authors in  \cite{BemaniAFDM_conf1} evaluated the performance of AFDM over high-mobility scenarios and in   high-frequency   bands. It is shown that AFDM is robust to the impairments such as  carrier frequency offset  and phase noise.  The time and frequency offset estimations for AFDM synchronization
  have  been studied in \cite{tang2023time}. 
 In \cite{YinChannel}, both single pilot and multiple pilots     aided channel estimation schemes for AFDM  were considered. In \cite{yin2022channel}, an embedded pilot-aided channel estimation scheme for AFDM with multiple input and multiple output (MIMO), called MIMO-AFDM, was investigated.  Moreover,  a  low complexity iterative decision feedback equalizer based on  weighted maximal ratio combining scheme  was introduced in \cite{bemani2022low}   for efficient symbol detection by exploiting the channel sparsity in AFDM.   A generalized message passing algorithm (MPA)-based detection scheme was developed in \cite{wu2023message}. However, the complexity of the MPA   detector  grows exponentially as the number of multipaths increases.  {By fully exploiting its }   inherent chirp nature of AFDM, it has also been applied for  integrated sensing and communications (ISAC). In particular, a low complexity algorithm based on matched filtering was developed in \cite{zhu2023low} for estimating high resolution target range with the AFDM waveform. In \cite{ni2022afdm}, an AFDM-based ISAC system was introduced, which utilizes a time-domain parameter estimation scheme.  Very   recently,  \cite{IntegratedAFDM} proposed an AFDM-based ISAC system using only chirp subcarrier. It shows that  sensing with one chirp  subcarrier allows   simple self-interference cancellation,
thus avoiding the need for expensive full duplex methods.

AFDM has also been integrated with code-domain non-orthogonal multiple access or modulation to achieve higher spectrum efficiency and enhanced reliability \cite{luo2023afdm,tao2023affine,zhu2023design}. Specifically,  \cite{luo2023afdm} and \cite{10683521} studied the AFDM-empowered sparse code multiple access  (SCMA) for massive connectivity over high mobility communications. In comparison to OFDM-SCMA, AFDM-SCMA exhibits significantly improved error performance.  In addition,  \cite{tao2023affine,zhu2023design} proposed index modulation (IM)-empowered AFDM systems, which was shown to outperform its OFDM-IM counterpart.

         \vspace{-1em}
\subsection{Motivations and Contributions}

The aforementioned works have shown  the   advantages of AFDM in high-mobility communications due to its ability to achieve \textit{full diversity}  \cite{AFDMTWCBemani,BemaniAFDM_conf1,luo2023afdm,tao2023affine,zhu2023design}.  However,  the \textit{full diversity} can only be   attained with a sophisticated receiver that implements      the optimal maximum \textit{a posteriori} (MAP) detection \cite{MaximumdIVERSITY}, such as MPA and sphere decoder.  In addition, it should be noted that existing works on AFDM generally focused  on uncoded systems and  {little has }   been reported on the coded AFDM system performances. Recently,  \cite{OCDMPerformance} compared the  uncoded and coded BER performances of OCDM and OFDM by employing a minimum mean square error (MMSE)    detector and  a rate-$1/3$ convolutional  code.  It was shown in  \cite{OCDMPerformance}  that   OCDM achieves  better uncoded BER performance but  similar coded BER performance compared to  OFDM. This is reasonable  due to the frequency diversity attained by coded OFDM.  The above observations can   be readily extended to AFDM systems due to its generic chirp nature.  
 {This also implies that enhanced }  detection and decoding in  AFDM should be carefully designed by considering its inherent features  for fully exploiting the  diversity gain.   {In addition, while there is extensive literature on efficient iterative receiver design, such as the Turbo iterative structure \cite{EnhancingLUo,hanzo2007turbo} and orthogonal approximate message passing design \cite{liu2024capacity,luo2023afdm}, developing more advanced and low-complexity  {AFDM receiver remains largely unexplored.} }

On the other hand,  the amalgamation  of  MIMO and AFDM, i.e., MIMO-AFDM,   can be leveraged to achieve higher spectrum efficiency  and lower error rate.  Although the channel estimation of MIMO-AFDM was studied in   \cite{YinChannel},    the efficient receiver design in MIMO-AFDM is a largely open research topic. 
 
Against the above background and research gaps, this paper aims to develop novel   low-complexity joint detection and decoding  schemes fully exploiting the system diversity in MIMO-AFDM systems.  Specifically, we  formulate  the   MIMO-AFDM  detection     as a Bayesian inference problem for  computing the marginal posterior distribution \cite{Zhaoofts}.  { We advocate the use of variational  inference (VI) \cite{Advancesvi} as an approximate inference algorithm in probabilistic models.}  The principle of VI is  to approximate the target posterior distribution using a simple distribution, thereby transforming an  intractable inference problem into a tractable one \cite{VIchen,WeijieVI}. Our second key innovation is to  represent  the joint distribution of  the transmitted  and received signal in MIMO-AFDM as a sparse factor graph, referred to as the VI graph. By adopting   low-density parity check  (LDPC)  codes   as the channel code \cite{WenJSG},  we merge the VI graph and  the relevant LDPC factor graph  into a joint sparse graph
(JSG). 

The main contributions of this paper are summarized as follows:  
\begin{itemize}
\item We    formulate MIMO-AFDM detection  as a Bayesian inference problem for computing the marginal posterior distribution. Specifically, we introduce a unified VI approach   to approximate the target posterior distribution, upon which  the  belief propagation (BP) and expectation propagation (EP)-based algorithms can be   derived.   
 Subsequently,  both  BP-JSG and EP-JSG are employed     for highly efficient message propagation.
\item Inspired by the  linear constellation encoding model, we introduce an  enhanced JSG (E-JSG)   to eliminate  the use of interleavers, de-interleavers, symbol-to-bit, and bit-to-symbol  log-likelihood ratio (LLR) transformations. As such, one can merge the detection and decoding for E-JSG by efficiently passing the belief messages over the integrated  sparse graph.  In addition,  for reduced receiver processing, we propose a   sparse channel method. The main idea of this approach is to  {approximate multiple graph edges} with insignificant channel coefficients into a single edge on the VI graph. 
\item We present extensive  numerical results to demonstrate   the superiority of the proposed JSG receivers   in terms of   lower computational complexity and reduced detection and decoding latency compared to conventional MMSE and turbo-like receivers. Furthermore, the proposed JSG receivers can  exploit multipath diversity and Doppler diversity in MIMO-AFDM systems more efficiently, giving rise to    improved coded BER performance compared to conventional receivers. 
 \end{itemize}
 
         \vspace{-0.5em}
\subsection{Organization}

This paper is organized as follows: Section \ref{Sysmodel} describes the system model of MIMO-AFDM.   The proposed VI-JSG is presented in Section \ref{SecJSG}, covering the principles of  VI, joint factor representation of MIMO-AFDM, and the message passing in VI-JSG.  Section \ref{EJSG} introduces the proposed  enhanced VI-JSG and the complexity reduction in the JSG. The simulation results are presented and discussed 
in Section \ref{Sim}. Section \ref{conclu}  concludes the paper.

         \vspace{-1em}
\subsection{Notation}

The $n$-dimensional complex, real and binary vector spaces are denoted as $\mathbb{C}^n$, $\mathbb{R}^n$ and $\mathbb{B}^n$, respectively.  Similarly, $\mathbb{C}^{k\times n}$, $\mathbb{R}^{k\times n}$ and $\mathbb{B}^{k\times n}$  denote the $(k\times n)$-dimensional complex, real and binary  matrix spaces, respectively.  
  {$\mathcal {CN}(0,1)$ denotes the
 complex Gaussian distribution with zero-mean and unit-variance.}  $ {{\mathbf{I}}_{n}}$ denotes an $n \times n $-dimensional  identity matrix.   $\text{diag}(\mathbf{x})$ gives a diagonal matrix with the diagonal vector of $\mathbf{x}$. $(\cdot)^\mathcal T$  and $(\cdot)^\mathcal H$ denote the transpose  and the Hermitian transpose operation, respectively.

\section{MIMO-AFDM Communication Model}
\label{Sysmodel}
This section introduces the MIMO-AFDM signal model. Specifically, we start by introducing the AFDM modulation, channel,  demodulation and input-output (I/O) relation in the single-input and single-output AFDM  (SISO-AFDM)  system, which is shown in Fig. \ref{SISO-AFDM}.    Subsequently, we extend the SISO-AFDM to the MIMO-AFDM case.





 \subsection{ SISO-AFDM Communication Model} 
\begin{figure}
    \centering
\includegraphics[width=0.8\linewidth]{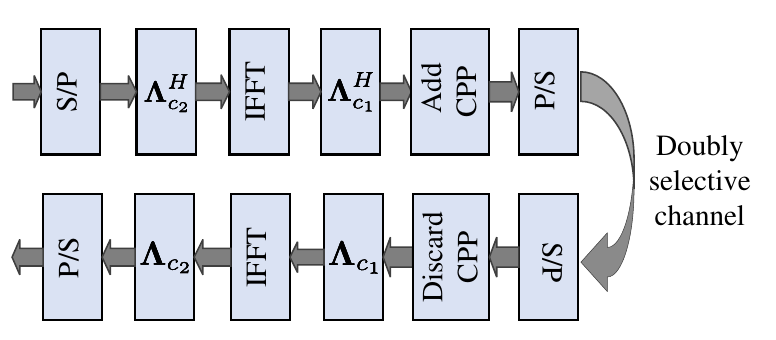}
    \caption{ {Example of SISO-AFDM communication model.}}
    \label{SISO-AFDM}
    \vspace{-1em}
\end{figure}

 {This section presents the signal model of a SISO-AFDM system, as shown in  Fig. \ref{SISO-AFDM}.}
 
 \subsubsection{AFDM modulation}  Denoted   { $ \mathbf x_{\text{SISO}} \in \mathbb C^{N \times 1}$} by the transmitted vector in the DAFT domain. Then     the modulated symbol is obtained by taking the  IDAFT of   { $ \mathbf x_{\text{SISO}} $},  i.e.,   
\begin{equation}
\small
\label{AFDMMOD}
    s_n = \sum_{m =0}^{N-1} x_m \varphi_{n}(m), n=0,1,\ldots,N-1,
\end{equation}
   where $\varphi_{n}(m)$ is the AFT  kernel at the $m$th SC and associated to the $n$th modulated symbol.   The AFT  kernel of   Type-A  in \cite{AFDMTWCBemani} is  considered, i.e., $\varphi_{n}(m)={1\over \sqrt{N}}e^{j2\pi(c_{1}n^{2}+c_{2}m^{2}+{nm\over N})},$
where $ c_1 \geq 0$ and $ c_2 \geq 0$ are the AFDM parameters. Note that (\ref{AFDMMOD}) can be written in the matrix form as
$\mathbf s = \Lambda_{c_1}^{\mathcal H} \mathbf F^{\mathcal H} \Lambda_{c_2}^{\mathcal H}  \mathbf x_{\text{SISO}} \equiv \mathbf A^{\mathcal H} \mathbf x_{\text{SISO}},$
 where $\mathbf A \equiv \Lambda_{c_2}  \mathbf F \Lambda_{c_1}  $ is the DAFT matrix,  $\Lambda_{c} =  \text{diag}\left ( e^{-j2\pi cn^2}, n=0,1, \ldots, N-1 \right)$,   $\mathbf F$ is the  DFT  matrix with its element at the $m$th column and the $n$th row given by   $e^{-j2\pi mn/N} / \sqrt{N}$. 

Then, a chirp-periodic prefix (CPP)  is applied to the transmitted signal \cite{AFDMTWCBemani}. The CCP with length $N_{\text{CPP}}$  is given by
  \begin{equation}
  \small
    s_n =   s_{N+n} e^{-j 2 \pi c_1(N^2+2Nn)}, n=-N_{\text{CPP}}, \cdots,-1.
\end{equation}

 \subsubsection{Channel} We consider a  doubly selective channel  with the channel response at time $n$ and delay $d$   given by    
\begin{equation}
\small
\label{channel}
    g_{n,d}  = \sum_{p=1}^{P} h_p e^{-j\frac{2\pi}{N} \nu_p n} \delta(d-d_p),
\end{equation}   
where $\delta(\cdot)$ denotes the Dirac delta function, $P$ is the number of paths, and $h_p$, $\nu_p$ and $d_p$ are the channel gain, Doppler shift  and the integer delay of the $p$th path, respectively. Note that $\nu_p$ is  normalized  with respect to the SC spacing and can be expressed as $\nu_p = \alpha_p +\beta_p$, where $\alpha_p \in [-\alpha_{\max},  \alpha_{\max}]$ and $\beta_p \in (-\frac{1}{2}, \frac{1}{2}]$ denote the integer and fractional parts of $\nu_i$, respectively,  and $ \alpha_{\max}$ is the maximum integer Doppler.

  \subsubsection{AFDM demodulation}    The received signal in the time domain after discarding the CPP can be expressed as  
\begin{equation}
\small
\label{rxSig}
    \widetilde{r}_n = \sum_{d=0}^{\infty} s_{n-d}g_{n,d} + \widetilde w_n,
\end{equation}   
where $w_n \sim \mathcal{CN}(0, N_0)$  is the additive Gaussian noise. After discarding the CPP, (\ref{rxSig}) can be re-written in the matrix form as 
$\widetilde{  \mathbf r}  = \sum_{p=1}^{P} {\widetilde{{\mathbf H}}_p} \mathbf s+ \widetilde{\mathbf w},$ 
 where $\widetilde{\mathbf w} $ is the noise vector in the time domain, ${\widetilde{{\mathbf H}}_p} =  h_p \mathbf{\Gamma}_{\text{CPP}_p} \mathbf{\Delta}_{\nu_p} \mathbf \Pi^{d_p}$ is the  time domain channel matrix of the $p$th path,   $\Delta_{\nu_p} = \text{diag}(e^{-j \frac{2\pi}{N} \nu_p  }, n=0,1,\ldots,N-1)$ models the Doppler effect, 
\renewcommand{\arraystretch}{0.4}
$   \mathbf  \Pi = {\left[ {\begin{array}{ cccc} 0& \cdots &0&1 \\ 1& \cdots &0&0 \\ \vdots & \ddots & \ddots & \vdots \\ 0& \cdots &1&0 \end{array}} \right]_{N \times N}} $ denotes the forward cyclic-shift matrix,   and 
  $\mathbf{\Gamma}_{\text{CPP}_p}    = \text{diag} \left( \left\{\begin{matrix}
  e^{-j2\pi c_1 (N^2-2N(d_p-n))}, &   n< d_p, \\ 
  1, & n \geq d_p,
 \end{matrix}\right. \right)$  denotes the effective CCP matrix.   
 
Finally, the received signal in the DAFT domain is obtained by applying the DAFT transform, i.e.,
\begin{equation}
\small
\label{DAFT}
 \mathbf r =  \Lambda_{c_2}  \mathbf F  \Lambda_{c_1}   \mathbf {\widetilde r }  \equiv \mathbf A \mathbf  {\widetilde r }.
\end{equation}

\subsubsection{The I/O relation}  Substituting (\ref{AFDMMOD}), (\ref{channel})  and  (\ref{rxSig}) into (\ref{DAFT}), yields the I/O relation in the time domain as 
\begin{equation}
\small
\begin{aligned}
     r_n = \frac{1}{N} \sum_{m = 0}^{N-1} \sum_{p = 1}^{P} & h_p  \eta(d_p,n,m)  \gamma(d_p,\nu_p,n,m)  x_m +w_n,
\end{aligned}
\end{equation}
where  
 \begin{subequations}
 \small
 \begin{align}
&\eta(d_p,n,m)  = {e^{j\frac{{2\pi }}{N}\left( {N{c_1}d_p^2 - m{d_p} + N{c_2}\left( {{m^2} - {n^2}} \right)} \right)}}, \\
& \gamma(d_p,\nu_p,n,m) =\frac{e^{-j2\pi (n-m+\text{Ind}_p + \beta_p)}-1}{e^{\frac{-j2\pi}{N} (n-m+\text{Ind}_p + \beta_p)}-1}, \label{indp}
\end{align}
 \end{subequations}
and $\text{Ind}_p = (\alpha_p + 2N c_1 d_p)_N$.   Note that the  I/O relation  in the DAFT domain can also be written in the matrix form, i.e.,   
\begin{equation}
\small
\begin{aligned}{\mathbf{r}} 
\label{inOut}
 & = \sum\limits_{p = 1}^P {{h_p}} \underbrace{{{\mathbf{\Lambda }}_{{c_2}}}{\mathbf{F}}{{\mathbf{\Lambda }}_{{c_1}}}{{\mathbf{\Gamma }}_{{\mathbf{CP}}{{\mathbf{P}}_p}}}{{\mathbf{\Delta }}_{{f_p}}}{{\mathbf{\Pi }}^{{d_p}}}{\mathbf{\Lambda }}_{{c_1}}^H{{\mathbf{F}}^H}{\mathbf{\Lambda }}_{{c_2}}^H}_{\overline{{\mathbf H}}_p}  {\mathbf{x}_{\text{SISO}}} + {\mathbf{w}_{\text{SISO}}} \\ & = {{\mathbf{H}_{\text{SISO}}}}   {\mathbf{x}_{\text{SISO}}} + {\mathbf{w}_{\text{SISO}}}, 
\end{aligned}
\end{equation}
 where ${\mathbf{H}_\text{SISO}}=\sum\nolimits_{p = 1}^P {{h_p}} {\overline{{\mathbf H}}_p}$ is the effective channel matrix and  {${\mathbf{w}_{\text{SISO}}}$} is the noise vector with the same distribution of ${\widetilde{\mathbf{ w}}}$.  It can be shown that the  element of  $ {\overline{{\mathbf H}}_p}$ at the $n$th row and the $m $th column  is 
 \begin{equation}
 \small
 \label{Hp}
 \begin{array}{l}  {\overline{{\mathbf H}}_p}[n,m] = \eta(d_p,n,m)  \gamma(d_p,\nu_p,n,m). \end{array} 
 \end{equation}

\begin{figure}
	\centering
	\begin{subfigure}{0.48\textwidth}
  \includegraphics[width=1.  \textwidth]{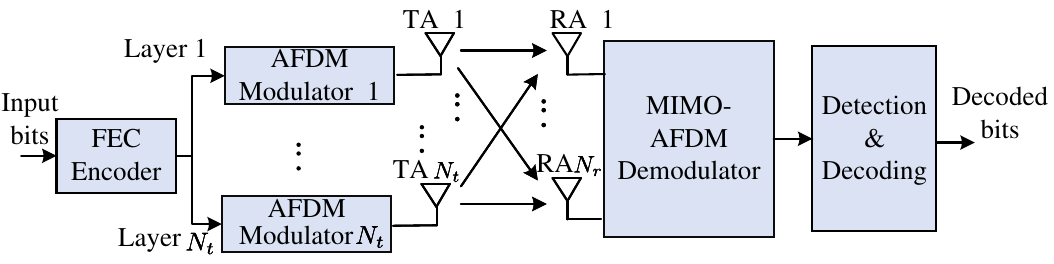}
		\caption{MIMO-AFDM of Scenario 1. }
	\end{subfigure}
	\begin{subfigure}{0.48\textwidth}
  \includegraphics[width= 1. \textwidth]{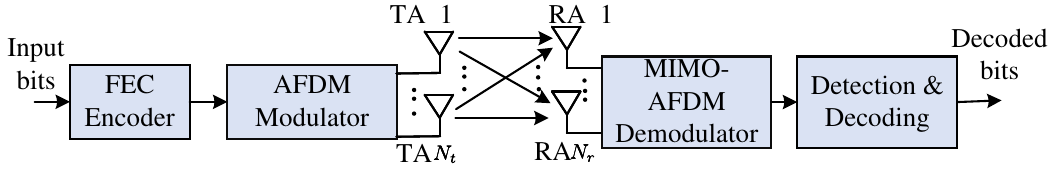}
		\caption{MIMO-AFDM of Scenario 2. }
	\end{subfigure}
	\caption{ {System model of MIMO-AFDM.}}
	\label{H_EFF_FACTOR}
         \vspace{-1.5em}
\end{figure}

In AFDM systems, the parameters $c_1$ and $c_2$ can be adjusted so that the non-zero elements of matrix $ {\overline{{\mathbf H}}_p}$ in each path do not overlap within ${{\mathbf{H}}_\text{SISO}  }$, resulting in a comprehensive delay-Doppler channel representation. 

 \vspace{-1em}
\subsection{Frameworks of MIMO-AFDM Schemes}

We consider a MIMO system where the  transmitter is equipped with $N_t$ transmit antennas (TAs) and the BS is equipped with $N_r$ receive antennas (RAs). Denoted  $\mathbf H^{(r,t)} \in \mathbb C^{N \times N} $ by the effective channel matrix from the $t$th TA to the $r$th RA.  Then the  received signal at the $r$th RA can be expressed as 
\begin{equation}
\label{SigMod1}
\small
    \mathbf y_{r} = \sum_{t =1}^{N_t}  \mathbf H^{(r,t)} \mathbf x_{t } +  \mathbf{w}_{r},
\end{equation}
where $\mathbf x_{t } \in \mathbb C^{N \times 1}$ denotes the transmitted data at the $t$th TA and $\mathbf{w}_{r} \in \mathbb C^{N \times 1}$ denotes the noise vector with $\mathcal{CN}(0, N_0)$ entries.  By staking the received signal of $N_r$ RAs, we have 
\begin{equation}
    \mathbf y  = \mathbf H\mathbf x +  \mathbf{w},
\end{equation}
where $  \mathbf y\equiv  \left [  \mathbf y_{1}^{\mathcal T}, \mathbf y_{2}^{\mathcal T}, \ldots,  \mathbf y_{N_r}^{\mathcal T} \right ]^{\mathcal T} \in \mathbb C^{NN_r\times 1 }$, $  \mathbf x \equiv \left [  \mathbf x_{1}^{\mathcal T}, \mathbf x_{2}^{\mathcal T}, \ldots,  \mathbf x_{N_t}^{\mathcal T} \right ]^{\mathcal T} \in \mathbb C^{NN_t\times 1 }$, $  \mathbf w  \equiv \left [  \mathbf w_{1}^{\mathcal T}, \mathbf w_{2}^{\mathcal T}, \ldots,  \mathbf w_{N_r}^{\mathcal T} \right ]^{\mathcal T} \in \mathbb C^{NN_r\times 1 }$, and 
\begin{equation}
\small
\begin{aligned}
    \mathbf H \equiv \left [ { \begin{matrix}   \mathbf H^{(1,1)} &  \mathbf H^{(1,2)} & \cdots  & \mathbf H^{(1,N_t)}\\
      \mathbf H^{(2,1)} &  \mathbf H^{(2,2)} & \cdots  &\mathbf H^{(2,N_t)} \\
       \vdots & \vdots &  \ddots & \vdots \\
             \mathbf H^{(N_r,1)} &  \mathbf H^{(N_r,2)} & \cdots  &\mathbf H^{(N_r,N_t)} \end{matrix} }\right ] \in \mathbb C^{NN_r\times NN_t }.
\end{aligned}
\end{equation}


Considering the relationship between signals transmitted on different TAs  and resource utilization strategies, the MIMO-AFDM frameworks can be categorized into the following two scenarios with the signal mode shown in Fig.   \ref{H_EFF_FACTOR}.

1) Scenario 1 (Multiplexing): As shown in \ref{H_EFF_FACTOR}(a), the coded bits are split into $N_t$ streams. the $n_t$th stream is      then modulated before  transmission over the  $n_t$th antenna.

2) Scenario 2 (Diversity): As shown in \ref{H_EFF_FACTOR}(b), the coded bits   are directly modulated with an AFDM modulator. All the $N_t$  TAs transmit the same modulated signal for diversity enhancement.

\section{ Variational Inference and Graphical Model}
\label{SecJSG}

 {In this section, we first present the VI framework \cite{Advancesvi} and the proposed   JSG representation in MIMO-AFDM. Then, we detail the message propagation process over the JSG based on VI principles.}

\subsection{Variational Inference}
Given   the received observation $\mathbf y$, the goal is to   
  infer the transmitted signal $\mathbf x$.   The optimum detection is MAP, which is given by 
  \begin{equation}
  \small
  \label{MAP}
  \hat {\mathbf{x}} = \mathop {\mathrm {\arg \!\max }} _{{  x_m\in {\mathcal{ X}, \forall m}}}  \quad p ({\mathbf{x}}\,|\, {\mathbf{y}}),
  \end{equation}
where $\mathcal{ X}$  denotes the constellation alphabet. Solving (\ref{MAP}) requires a computational complexity order of   $\vert\mathcal X \vert^{NN_t} $ and  $\vert\mathcal X \vert^{N} $  for Scenarios 1 and   2, respectively, which   increases   exponentially with  $N$ or $N_t$.  The joint probability of the transmitted vector $\mathbf x$ and received signal $\mathbf y$ can be expressed as: $p\left ({\mathbf {x},\mathbf {y}}\right)=p\left ({\mathbf {x}}\right)p\left ({\mathbf {y}|\mathbf {x}}\right)$, where $p\left ({\mathbf {x}}\right)$ denotes the   prior probability of $\mathbf x$ and $p\left ({\mathbf {y}|\mathbf {x}}\right)$ is the  likelihood probability. From Bayes’ rule, the \textit{a posterior} probability of  $\mathbf x$ is  { given by \cite{Advancesvi}}
 \begin{equation}
 \label{prost}
 p\left ({\mathbf {x}|\mathbf {y}}\right)=\frac {p\left ({\mathbf {x}}\right)p\left ({\mathbf {y}|\mathbf {x}}\right)}{p\left ({\mathbf {y}}\right)}\propto p\left ({\mathbf {x}}\right)p\left ({\mathbf {y}|\mathbf {x}}\right). 
 \end{equation}
  The marginal posterior probability is calculated as $p\left ({ {x}_{m}|\mathbf {y}}\right)=\sum _{\sim  {x}_{m}}p\left ({\mathbf {x}|\mathbf {y}}\right)$  which serves as the soft information output for the channel decoder,  where $\sum _{\sim  {x}_{m}}$ denotes the  summation over all the arguments except  $x_m$.  Unfortunately, the direct calculation of  $p\left ({ {x}_{m}|\mathbf {y}}\right)$ incurs  high computational complexity.  As an approximation  approach,  the VI  technique can be employed for  {efficient compute}  (\ref{prost}).    The  { core idea }  of VI is to find a distribution $q(\mathbf x \vert \mathbf y)$ from a tractable distribution family $Q$, serving as an optimized approximation of the target posterior distribution   $p(\mathbf x \vert \mathbf y)$. As such, the inference problem in (\ref{prost}) can be transformed to a tractable optimization problem.  
The widely used optimization metrics in VI are the exclusive Kullback-Leibler (KL) divergence $\text{KL} ( p \vert q)$ and inclusive KL divergence $\text{KL} ( q \vert p) $, which measure the similarity between two distributions. The $\text{KL} ( p \vert q)$  and $\text{KL} ( q \vert p) $ are respectively defined as \cite{Advancesvi}
\begin{equation}
\small
\begin{aligned} \text{KL} ( q \vert p) =&\sum _{\mathbf {x}}q\left ({\mathbf {x}|\mathbf {y}}\right)\log \frac {q\left ({\mathbf {x}|\mathbf {y}}\right)}{p\left ({\mathbf {x}|\mathbf {y}}\right)},  \\ 
\text{KL} (p\vert q)=&\sum _{\mathbf {x}}p\left ({\mathbf {x}|\mathbf {y}}\right)\log \frac {p\left ({\mathbf {x}|\mathbf {y}}\right)}{q\left ({\mathbf {x}|\mathbf {y}}\right)}.  
\end{aligned}
\end{equation}
 
Note that the KL divergence is non-negative. 
 Moreover, it is neither symmetric nor a distance metric, i.e., $\text{KL} ( p \vert q) \neq \text{KL} ( q \vert p)$.    $ \text{KL} ( q \vert p)$ divergence   emphasizes  assignment of low probability mass of $q$ to the location where $p$ is very small. In other words, $ \text{KL} ( q \vert p)$  has the propriety of  ``mode-seeking" or   ``zero-forcing" as it   forces $q$  to concentrate on one of the local maximum values of $p$. Conversely,  $\text{KL} (p\vert q)$   divergence   emphasizes  assignment of high probability mass of $q$ to the location where $p$ has positive mass. Namely,  $\text{KL} (p\vert q)$ metric has the property of mean-seeking or  mass covering, meaning  that   $q$ tends to fit the mean value of  $p$.  

In a nutshell, the VI framework provides a unified structure to design detection algorithms by taking $\text{KL} ( q \vert p)$ and $\text{KL} (p\vert q)$ into consideration. In addition, the  VI framework  can also be described  by using a powerful probabilistic graphical model  known as a factor graph. This graphical representation is the key for low-complexity  algorithm design, which will be discussed in the sequel.

\begin{figure}
	\centering
	\begin{subfigure}{0.5\textwidth}
 	\centering
  \includegraphics[width = 0.5  \textwidth]{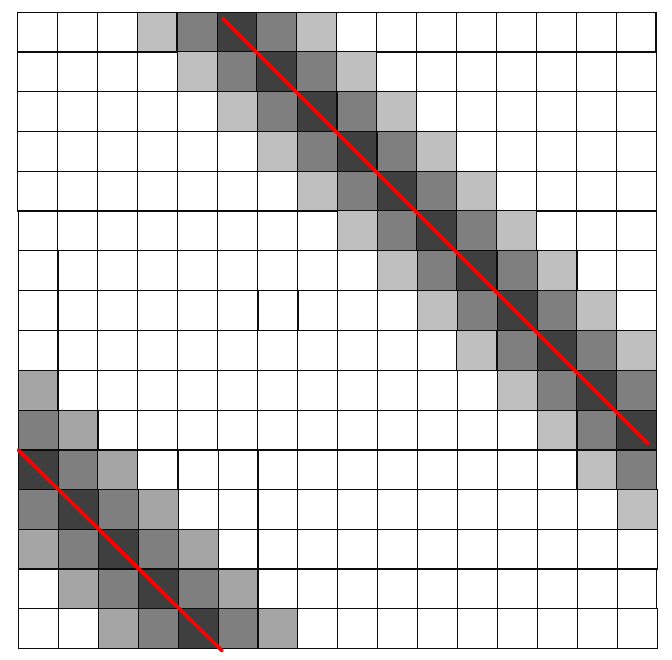}
		\caption{An example of the channel matrix $\mathbf H_p$ with $k_{\nu}=2$. }
	\end{subfigure}
	\begin{subfigure}{0.45\textwidth}
  \includegraphics[width= 1  \textwidth]{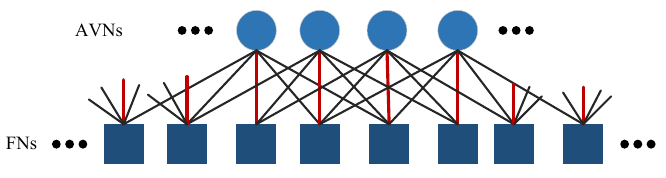}
		\caption{The corresponding factor representation of $\mathbf H_p$ }
	\end{subfigure}
	\caption{An example of a channel matrix and its corresponding factor graph representation.}
	\label{Hpfactor}    
 \vspace{-2em}
\end{figure}

     \vspace{-1em}
\subsection{Joint Factor Representation in MIMO-AFDM}
\label{factorGraph}
 
Recall (\ref{Hp}),  for each path, $\eta(d_p,n,m) $ has unit  energy, and one has   \cite{AFDMTWCBemani}
\begin{equation}
\small 
\begin{aligned}
    \left  \vert {\overline{\mathbf{H}}_p}[n,m]  \right   \vert  & = \left \vert\frac{1}{N} \gamma(d_p,\nu_p,n,m) \right  \vert = \left \vert \frac{\sin{(N\theta)}}{N\sin{(\theta)}}   \right  \vert \\
     & \underset{\leq}{(\text{i})} \frac {N-1}{N}\left |{{\cos (\theta)}}\right |+\frac {1}{N},
\end{aligned}
\end{equation}
where  $\theta \triangleq \frac {\pi }{N}(n-m + \mathrm {Ind}_{p}+\beta_{p})$, $ {\overline{\mathbf{H}}_p}[n,m]$  denotes the entry at the $n$th row and $m$th column of  ${\overline{\mathbf{H}}_p}$,
and the detailed derivation of $(\text{i})$ is given by  the Equations (35) and (36) of \cite{AFDMTWCBemani}. The above bounds hold tight as $N$ increases.    It is noted that 
$\gamma(d_p,\nu_p,n,m)$ achieves the peak energy at $m = (n+\text{Ind}_p)_N$, and    the elements  in  ${\overline{\mathbf{H}}_p}[n,m]$ 
 become insignificant as $m$ moves away from  $(n+\text{Ind}_p)_N$. 
   To better present the message passing over the VI graph, we first assume that those  insignificant elements in $ {\overline{\mathbf{H}}_p}[n,m]$  are  zeros when $m$  falls outside of $[n+\text{Ind}_{p}-\nu_v, n+\text{Ind}_{p}+\nu_v]$.       Namely, 
\begin{equation}
\small
\label{Hpath}
\begin{aligned}
   &    {\overline{\mathbf{H}}_p}[n,m]       = \left\{\begin{matrix}
   {\overline{\mathbf{H}}_p}[n,m],       & \makecell { {{\left( {n + {{\operatorname{Ind} }_p}} - k_{\nu}\right)}_N} \leq m \\    \leq {{\left( {n + {{\operatorname{Ind} }_p}} + k_{\nu}\right)}_N}} \\ 
 0, & {{\text{ otherwise }}}
\end{matrix}\right.
\end{aligned}.
\end{equation}
In Subsection  \ref{Sparseejsg}, we will present the proposed sparse channel scheme to address  these insignificant elements and determine $k_{\nu}$. 

\begin{figure}
     \centering
     \includegraphics[width=0.9\linewidth]{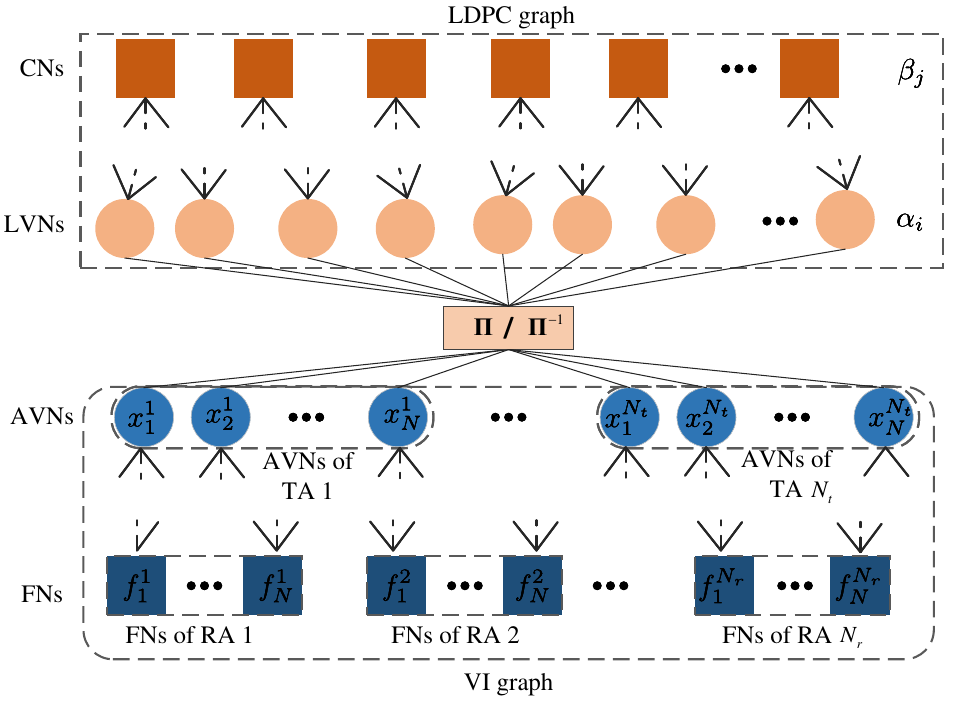}
     \caption{ {Factor graph representation of the proposed JSG.}}
     \label{JSG_FACTOR}   
     \vspace{-2em}
 \end{figure}

\color{black}
A factor graph is a bipartite graph representing the factorization of a  joint distribution function $p(\mathbf x, \mathbf y)$. Fig. \ref{Hpfactor}(a) shows an example of  ${\overline{\mathbf{H}}_p}$ in a SISO-AFDM system with $N=16$ and $k_{\nu}=2$.    Consider the one-path channel in Fig. \ref{Hpfactor}(a),   the   factor graph is shown in   Fig. \ref{Hpfactor}(b).  Each  AFDM variable node (AVN) is connected to $2k_{\nu}+1$  function nodes (FNs) and each FN  is also  occupied by $2k_{\nu}+1$ AVNs. For simplicity, denote $ x_{m}^{t}$ as the transmitted symbol at the $m$th SC of the $t$th TA.  Consider the MIMO-AFDM of  Scenario $1$, the  joint distribution $ p(\mathbf x, \mathbf y)$ can be  factorized as
\begin{equation}
\small
\begin{aligned}
   p(\mathbf y,\mathbf x)= &p(\mathbf y \vert \mathbf x)p(\mathbf x) \\=&  \prod _{t=1}^{N_t} \prod _{m=1}^{N} p( x_{m}^{t})  \prod _{r=1}^{N_{r}}\prod _{n=1}^{N} p(y^{r}_{n}\vert \mathbf x),  
\end{aligned}
\label{pxy}
\end{equation}
where $p\left ({y_{n}^{r}|\mathbf {x}}\right)=\mathit {\mathcal {CN}}\left({y_{n}^{r};\sum _{t}\sum _{m }h_{n,m}^{r,t}x_{m}^{t}, N_0}\right)$, $h_{n,m}^{r,t}$ denotes the channel coefficient from the  $m$th SC of the $t$th TA to the $n$th SC of the $r$th RA,   and $p( x_{m}^t)$ denotes the prior probability of $x_m^t$. The factor graph   can be designed to contain $NN_t$  AVNs  $x_m^t$ and $N_rN$ likelihood FNs $f_{n}^{r}$.  
The    LDPC codes can also be represented by  a bipartite graph called Tanner graph with the belief messages  passing  over the LDPC VNs (LVNs) and LDPC check nodes (CNs).  In this paper, we propose to design the JSG-based receiver   by merging the VI factor graph   and the LDPC graph into an integrated one. As a result, the belief messages  are exchanged iteratively on a JSG. Fig. \ref{JSG_FACTOR} shows an example of the JSG  for the MIMO-AFDM of Scenario $1$.  The AVNs and FNs represent the transmit layers and the likelihood function $p(y^{r}_{n}\vert \mathbf x)  $.  
The set of AVN  indices sharing the $n$th  FN of the $r$th RA is denoted as $F(n,r)=\left \{{ (m,t): h_{n,m}^{r,t} \neq 0}\right \}$, whereas  the set of FN  indices sharing the $m$th  VN is denoted by $V(m,t)=\left \{{n,r: h_{n,m}^{r,t} \neq 0}\right \}$.   In addition, the cardinalities of  $V(m,t)$ and  $F(n,r)$ are denoted as $\vert V(m,t) \vert = d_f$  and $\vert F(n,r) \vert = d_v$, respectively.  
In this paper, we first consider the MIMO-AFDM in  Scenario $1$, and the results can be readily extended to   Scenario $2$. 

 
\vspace{-0.4cm}
\subsection{BP-JSG}
In this subsection, we first consider the  optimization metric as the exclusive KL divergence, i.e., $\text{KL} ( p \vert q)$. %
It is well known that the $\log(\cdot)$  partition function   $ \log  p({\mathbf y})$ can be decomposed as \cite{Factorbp}
\begin{equation}
\small
\begin{aligned}
    \log & p({\mathbf y})     =\sum_{\mathbf x } q(\mathbf x) \log \left(  \frac{p(\mathbf x, \mathbf y) }{q(\mathbf x)}   \times  \frac{q(\mathbf x ) }{p(\mathbf x \vert \mathbf y)} \right) \\
& = \underbrace{\sum _{\mathbf {x}} q\left ({\mathbf {x} }\right)\log p\left ({\mathbf {x},\mathbf {y}}\right)-\sum _{\mathbf {x}}q\left ({\mathbf {x} }\right)\log q\left ({\mathbf {x} }\right) }_{\mathscr {L_{\mathrm {ELBO}}}}    +   \text{KL}  ( q \vert p),
\end{aligned}
\end{equation} 
where ${\mathscr {L_{\mathrm {ELBO}}}}$ is the evidence lower bound (ELBO). Namely,  ${\mathscr {L_{\mathrm {ELBO}}}} = \log  p({\mathbf y}) -  \text{KL}  ( q \vert p)$.  As a result, instead of minimizing  $\text{KL} ( p \vert q)$, one can  maximize ${\mathscr {L_{\mathrm {ELBO}}}}$. To solve this problem, we resort to the Bethe approximation  for  maximizing ${\mathscr {L_{\mathrm {ELBO}}}}$. It is shown that the stationary points of the Bethe approximation to the ${\mathscr {L_{\mathrm {ELBO}}}}$ are equivalent to the fixed points of the  {standard  BP algorithm \cite{Factorbp,VIchen,meng2015expectation}.}   Denote $   L_{v_{m}^{t}\rightarrow f_{n}^{r}}^{l}({x}_{m}^{t})$ and $ L_{f_{n}^{r} \rightarrow v_{m}^{t} }^{l}( {x}_{m}^{t}) $ as the LLR propagated from VN $v_{m}^{t}$  to  FN   $f_{n}^{r}$ and   FN  $f_{n}^{r}$  to  VN $v_{m}^{t}$at the $l$th iteration, respectively.  The iterative messages exchanged between  FNs and  AVNs are computed as
 \begin{equation}
  \small
  \label{FN_up}
 \begin{aligned}
L_{f_{n}^{r} \rightarrow v_{m}^{t} }^{l}( {x}_{m}^{t}) = &  \max_{ \forall {x}_{  m  \prime}^{ t \prime  },  {x}_{  m  \prime}^{ t \prime  } \neq  {x}_{  m  }^{ t   } }      - \frac{1}{2N_0} \left \vert y_k -     \sum\limits_{  {   m, t  \in F(n,r)}}h_{n,m}^{r,t}  x_{m}^t \right \vert ^2  \\
&  \quad + \sum_{ m, t  \in F(n,r)} L_{v_{m}^{t}\rightarrow f_{n}^{r}}^{t}({x}_{m}^{t}),
 \end{aligned}
\end{equation}
where 
\begin{equation}
 \small
  \label{VN_up}
L_{v_{m}^{t}\rightarrow f_{n}^{r}}^{t}({x}_{m}^{t}) =      \sum _{ n,r \in V(m,t)\backslash \lbrace n,r\rbrace } I_{f_{n}^{r} \rightarrow v_{m}^{t} }^{l-1}( {x}_{m}^{t})  , \end{equation}
where  $n,r \in V(m,t)\backslash \lbrace n,r\rbrace $ represents removing the  FN $f_n^r$ from the set $V(m,t)$.

The message propagation within the LDPC graph and between the LDPC and VI graph will be discussed along with the proposed EP-JSG.


     \vspace{-0.8em}

\subsection{EP-JSG}

In BP-JSG, (\ref{FN_up})-(\ref{VN_up}) require global searches over the joint space of all symbols, which is a discrete set. Consequently, the computational complexity of BP-JSG is exponential, i.e., $\mathcal O\left( M^{P k_{\nu}} \right)$.   As $M$, $P$, and $k_{\nu}$ increase, BP-JSG becomes impractical  in real-world systems, especially in scenarios with high mobility and rich scattering environments as $P$ and $k_{\nu}$   typically have  large values.  Instead of using $\text{KL} (q\vert p)$, we resort to the inclusive $\text{KL} (p\vert q)$ as the optimization metric. The  objective is to find a distribution family $q(\mathbf x)$ that minimizes $\text{KL} (p\vert q)$. Although global optimization is intractable,  {it can be iteratively solved using the well-known EP algorithm \cite{EPT,VIchen,meng2015expectation}.}  Denote $   I_{v_{m}^{t}\rightarrow f_{n}^{r}}^{l}({x}_{m}^{t})$ and $ I_{f_{n}^{r} \rightarrow v_{m}^{t} }^{l}( {x}_{m}^{t}) $ as the belief messages propagated from VN $v_{m}^{t}$  to  FN   $f_{n}^{r}$ and     $f_{n}^{r}$  to   $v_{m}^{t}$ at the $l$th iteration, respectively.   According to the EP principles, the message update rule can be expressed as \cite{EPT}
\begin{equation}
\small
\label{IV2F}
  I_{v_{m}^{t}\rightarrow f_{n}^{r}}^{l}({x}_{m}^{t})\propto\frac{\text{Proj}_{\Phi}(q_0^{~l}({x}_{m}^{t}) )} {I_{f_{n}^{r} \rightarrow v_{m}^{t} }^{l-1}( {x}_{m}^{t})}, 
\end{equation}
\begin{equation}
  I_{f_{n}^{r} \rightarrow v_{m}^{t} }^{l}( {x}_{m}^{t}) \propto \frac{\text{Proj}_{\Phi}(q_{n,{r}}^{t}({x}_{m}^{t}))}{ I_{v_{m}^{t}\rightarrow f_{n}^{r}}^{l}({x}_{m}^{t})},
\end{equation}
where $\text{Proj}_{\Phi}(p)=\arg\min\limits_{q\in\Phi} \text{KL}(p\vert q)$ denotes   the projection of a particular distribution $p$ into some distribution set $\phi$ for the given metric of $\text{KL}(p\vert q)$, and  
\begin{equation}
    q_0^{~l}({x}_{m}^{t})          \propto p_0({x}_{m}^{t}) \prod _{r=1}^{N_{r}} \prod _{ n,r \in V(m,t)} I_{f_{n}^{r} \rightarrow v_{m}^{t} }^{l}( {x}_{m}^{t}),  
\end{equation}
\begin{equation}
\small
\label{qnn}
   q_{n, {r}}^{t}({x}_{m}^{t})   \propto \sum_{  {x}_{ m\prime}^{ t\prime  } \neq {x}_{ m }^{ t }} p(y_{n}^{r} \vert {x}_{  m\prime}^{ t\prime}) \prod_{ m\prime, t\prime  \in F(n,r)\backslash \lbrace m,t \rbrace  } I_{v_{m}^{t}\rightarrow f_{n}^{r}}^{t}({x}_{m}^{t}).
\end{equation}
 The above procedure can be understood as the  messages propagation over the  VI  graph depicted in Fig. \ref{JSG_FACTOR}. 
It treats the messages exchanged between  VNs  and    FNs  as continuous random variables and approximates the true distribution with a Gaussian distribution, characterized by its mean and variance. This approach enables signal detection to be transformed into the computation of mean and variance, avoiding the need to traverse all symbols. Specifically, we have
\begin{subequations}
\small
\begin{align}
       \text{Proj}_{\Phi}(q_0^{~l}({x}_{m}^{t}) ) \propto & \mathcal {CN}({\mu _{m,t}^{l},\xi _{m,t}^{l}}),  \label{eq_a} \\ 
       I_{v_{m}^{t} \rightarrow  f_{n}^r}^{l}(x_m^t)\propto & \mathcal {CN}({\mu _{m,t\rightarrow  n,r}^{l},\xi _{m,t\rightarrow  n,r}^{l}}), \label{eq_b} \\ I_{ f_n^r\rightarrow v_m^t}^{l}(x_{m}^t) \propto & \mathcal {CN}({\mu _{n,r\rightarrow m,t}^{l},\xi _{ n,r\rightarrow m,t}^{t}}). \label{eq_c}
\end{align}
\end{subequations}
Now, we introduce the detailed  message propagation in the proposed JSG.

  \subsubsection{Updating of VNs} In the separate graph case,  VNs only gather information from one type of nodes (FNs or   CNs). However, in the JSG, the updating of VNs in JSG involves the information from both sub-graphs. 
  Let  $ L_{  \alpha_{i}  \rightarrow \beta_{j} }$ be $L_{\beta_{j} \rightarrow \alpha_{i   }}$ by the LLR updating from the  LVN  $\alpha_{i}$ to the CN  $ \beta_{j}$, and   the  $ \beta_{j}$ to the   $\alpha_{i}$, respectively.  Then, the updated $ L_{  \alpha_{i}  \rightarrow \beta_{j} }$ is given by
\begin{equation}
\small
\label{LVNUpd}
    L_{  \alpha_{i}  \rightarrow \beta_{j} }=  L^{a,{\mathrm {LDPC}}}_{\alpha_{i}}  + \sum_{j ^{\prime} \in \eta_i  \setminus j } L_{\beta_{j^{\prime}} \rightarrow \alpha_{i   }},
\end{equation}
where $L^{a,{\mathrm {LDPC}}}_{\alpha_{i}}$ denotes the \textit{a prior} LLR updated from VI graph at the LVN $\alpha_{i}$,  and $\eta_i$ denotes the set of CNs connected to LVN $\alpha_{i}$.  

Denote $ L^{a,{\mathrm {VI}}}(x_{m}^t)$ as the \textit{a prior} LLR  updated from LDPC graph to AVN  $x_{m}^t$.  The AVN nodes gather the  LLR information from both the LDPC graph  and FNs  as follows
\begin{equation}
\small
\label{LLR_xmt}
    Lq( x_{m}^t  \vert \mathbf y) =  L^{a,{\mathrm {VI}}}(x_{m}^t) + {\sum _{(n,r) \in V(m,t) }  L_{f_n^r\rightarrow v_m^t}^{l}( x_m^t)}.
\end{equation}
Then, based on  (\ref{LLR_xmt}), the \textit{a posterior} belief probability for  AVN $x_{m}^t$ can be calculated as
\begin{equation}
    \small
   p( x_{m}^t  \mid \mathbf y) = \frac {\exp \left (  Lq( x_{m}^t  \vert \mathbf y) \right)}{{1 + \exp \left (  Lq( x_{m}^t \vert \mathbf y) \right)}}. 
\end{equation}
Accordingly, the mean and variance of  AVN $x_{m}^t$ can be expressed as
\begin{equation}
\label{meanVar}
\small
\begin{aligned}
       \mu _{m,t}^{l} =&\sum _{ a_{m}^t\in \mathcal {X}   } p( x_{m}^t=  a_{m }^t \mid \mathbf y) a_{m}^t, \\ 
       \xi _{m,t}^{l} = &\sum _{  a_{m}^t \in \mathcal {X} } p( x_{m}^t=  a_{m}^t \mid \mathbf y){\left |{a_{m}^t-\mu _{m,t}^{l}}\right |}^{2}.
\end{aligned}
\end{equation}
Upon obtaining the  mean $\mu _{m,t}^{l}$ and variance $\xi _{m,t}^{l}$,  by substituting ({\ref{eq_a}})-({\ref{eq_c}}) into (\ref{IV2F}), one has the mean  and variance  passing from VNs to FNs respectively  as follows
\begin{equation}
\small
\label{MVUpd}
\begin{aligned}
        \xi _{m,t\rightarrow n,r}^{l}=&\left ({\frac {1}{\xi _{m,t}^{l} }-\frac {1}{\xi _{n,r\rightarrow m,t}^{l-1}}}\right)^{-1}, \\
        \mu _{m,t\rightarrow n,r}^{l}=& \xi _{m,t\rightarrow n,r}^{l} \left({\frac {\mu _{m,t}^{l} } { \xi _{m,t}^{l}}  -  \frac { \mu _{n,r\rightarrow m,t}^{l-1}} {\xi _{n,r\rightarrow m,t}^{l-1} 
 }}\right).
\end{aligned}
\end{equation}

\subsubsection{Updating of FNs and CNs}

 The LLR of the CN $\beta_j$ is updated as
\begin{equation}
\small
\label{CNupd1}
    L_{ \beta_{j}  \rightarrow \alpha_{i}}=\gamma^{-1}\left(\sum_{i^{\prime}\in\phi_{j} \setminus i}\gamma\left(L_{\alpha_{i^{\prime}} \rightarrow \beta_j}\right)\right),
\end{equation}
where  $\phi_{j} \setminus i$ is the set of VNs (excluding $\alpha_{i}$) that connect to the CN $\beta_{j}$, and $ \gamma(x) $ and $\gamma^{-1}(x)$ are respectively defined as 
\begin{equation}
\small
\label{CNupd2}
\begin{aligned}
     \gamma(x) &=\text{sign}(x)\times\left(-\log\tan{\vert x\vert\over 2}\right), \\
    \gamma^{-1}(x)& =(-1)^{\text{sign}(x)}\times\left(-\log\tan{\vert x\vert\over 2} \right),
\end{aligned}
\end{equation}
where $\text{sign}(x)$ denotes the the sign of $x$. 

On the FN side, the calculation of  $\mu _{n,r\rightarrow m,t}^{l}$ and  $\xi _{n,r\rightarrow m,t}^{l}$   plays a crucial role in the convergence of expectation propagation. The received signal $y_n^r$ is considered as the sum of transmitted signals from all layers plus the noise.  Namely, we have
\begin{equation}
\small
    x_{m}^t=\frac {1}{h_{n,m}^{r,t}}\Bigg ({y^{r}_{n}-\sum _{ \substack {i,j\in F(n,r), \\ i\neq m, j \neq t}} h_{n,i}^{r,j} x_i^j+\omega ^{r}_{n}}\Bigg).
\end{equation}
Then, based on  ({\ref{eq_a}})-({\ref{eq_c}}) and (\ref{qnn}), the message in the reverse  direction becomes
 \begin{equation}
\small
     \mu _{n,r\rightarrow m,t}^{l} =\frac {y_{n}^{r}-Z _{n,r\rightarrow m,t}^{l} }{h_{n,m}^{r,t}}, \; \xi _{n,r\rightarrow m,t}^{l}=\frac {N_0+ B_{n,r\rightarrow m,t}^{l}}{\left |h_{n,m}^{r,t}\right |^{2}},
\end{equation}
where 
\begin{equation}
\small
\begin{aligned}
    Z _{n,r\rightarrow m,t}^{l} =& \sum _{ \substack {i,j\in F(n,r), \\ i\neq m, j \neq t}} h_{n,i}^{r,j} \mu _{i,j\rightarrow n,r}^{l-1},  \\
    B_{n,r\rightarrow m,t}^{l}=& \sum _{ \substack {i,j\in F(n,r), \\ i\neq m, j \neq t}} \vert h_{n,i}^{r,j} \vert ^2\xi _{ i,j\rightarrow n,r }^{l-1}.
\end{aligned}
\end{equation}
 
Finally, the posterior probabilities of transmitted symbol in LLR domain is given by
\begin{equation}
\small
    L_{f_n^r\rightarrow v_m^t}^{l}( a_m^t) =  \frac{-\left |{a_m^t- \mu _{n,r\rightarrow m,t}^{l} } \right | ^{2}}  {\xi _{n,r\rightarrow m,t}^{l}},  \; \forall a_m^t \in \mathcal X.
\end{equation}


\subsubsection{Extrinsic and prior LLRs exchange}

The output  extrinsic information of the VI detector are  utilized as the \textit{a priori} LLRs for the LDPC decoder.  Specifically, the extrinsic information at the AVN $x_m^t$ is calculated as 
\begin{equation}
\small
    L^{e,VI}_{m,t}(a_m^t) = \sum_{r=1}^{N_r} \sum_{n=1}^{N}  L_{f_n^r\rightarrow v_m^t}^{l}( a_m^t).
\end{equation}
 Denote $ \mathbf c_{m,t} = \{c_{m,t}^i \vert i = 1,2, \ldots, \log_2M\} \in  \mathbb B^{\log_2M \times 1}$ as the corresponding bits of the symbol $a_m^t$. Then, $ L^{e,VI}_{m,t}(a_m^t)$ is further demapped to  the  bit-level  LLRs, which can be expressed as  
\begin{equation}
\small
\label{LLR1}
\begin{aligned}
    L^{e,VI}_{m,t}(c_{m,t}^i) =  &\mathrm {log}\frac {\sum _{\forall \mathbf c_{m,t} :c_{m,t}^i =0}  \exp(   L^{e,VI}_{m,t}(a_m^t) )} {\sum _{\forall \mathbf c_{m,t} :c_{m,t}^i =1} \exp(  L^{e,VI}_{m,t}(a_m^t))}. \\ 
\end{aligned}
\end{equation}
Let $L^{e,VI}_{m,t}(\mathbf c )$ denote  the extrinsic bit LLRs  from $N_t$ transmit antennas. $L^{e,VI}_{m,t}(\mathbf c )$  is further deinterleaved as the \textit{a priori} LLRs input to the  LDPC decoder, i.e.,
\begin{equation}
\small
    L^{a,{\mathrm {LDPC}}}(\boldsymbol{\alpha})  = {\Pi ^{ - 1}} \left (   L^{e,VI}_{m,t}(\mathbf c ) \right).
\end{equation}

At the LVN side, it  gathers information from CNs by  
\begin{equation}
\small
   {L} ^{e,\text{LDPC}}(\alpha_i) =   \sum_{j \in \eta_i}     L_{\beta_{j} \rightarrow \alpha_{i} }.
\end{equation}
After interleaving, the resultant LLR can be expressed as  
\begin{equation}
\small
    {L} ^{a,\text{VI}}(\mathbf{c}) = \Pi ({L} ^{e,\text{LDPC}}(\boldsymbol{\alpha})).
\end{equation}
To pass the belief messages from LVNs to AVNs, one needs to transform bit LLRs to symbol LLRs. Specifically,   the \textit{a priori } symbol-level LLRs can be calculated as 
 \begin{equation} 
\small
\begin{aligned}
 L^{a,{\mathrm {VI}}}(x_{m}^t)  & = \sum_{i=1}^{\log_2(M)} \log \frac { \exp \left ( - \Psi_{\mathcal X}[i]  {L} ^{a,\text{VI}}_{\boldsymbol{\alpha}}(c_{m,t}^i))   \right)}{{1 + \exp \left (  {L} ^{a,\text{VI}}_{\boldsymbol{\alpha}}(c_{m,t}^i))   \right)}},\\
 \end{aligned}
 \label{bit2sym}
\end{equation}
where  $\Psi_{\mathcal X}[i] \in \{0,1\}$ denotes the labeling value of the $i$th bit of $\mathcal X$.

\textit{Remark 1:  The proposed JSG allows belief messages from the LDPC
and VI graphs to propagate in a joint manner. In addition, the
AVNs (FNs) and LVNs (CNs) are able to update messages
simultaneously, leading to faster convergence with fewer iterations.
}

     \vspace{-0.8em}

\section{The proposed  E-JSG}
\label{EJSG}
Note that messages in the VI graph are updated at the symbol level, whereas messages in the LDPC graph are updated on a per-bit basis.  {This mismatch between symbol-level message updates in the VI graph and bit-level updates in the LDPC graph results in inefficient message propagation.} Furthermore, the computation of bit-to-symbol and symbol-to-bit LLRs introduces higher computational complexity.

 {To address these challenges, we propose an  E-JSG, which enables message updating on a per-bit basis.} Moreover, to further reduce computational complexity, we introduce a sparse channel method based on the  eSNR, { which helps eliminate unnecessary message propagation over the E-JSG.}

 \begin{figure}
     \centering
     \includegraphics[width=0.65\linewidth]{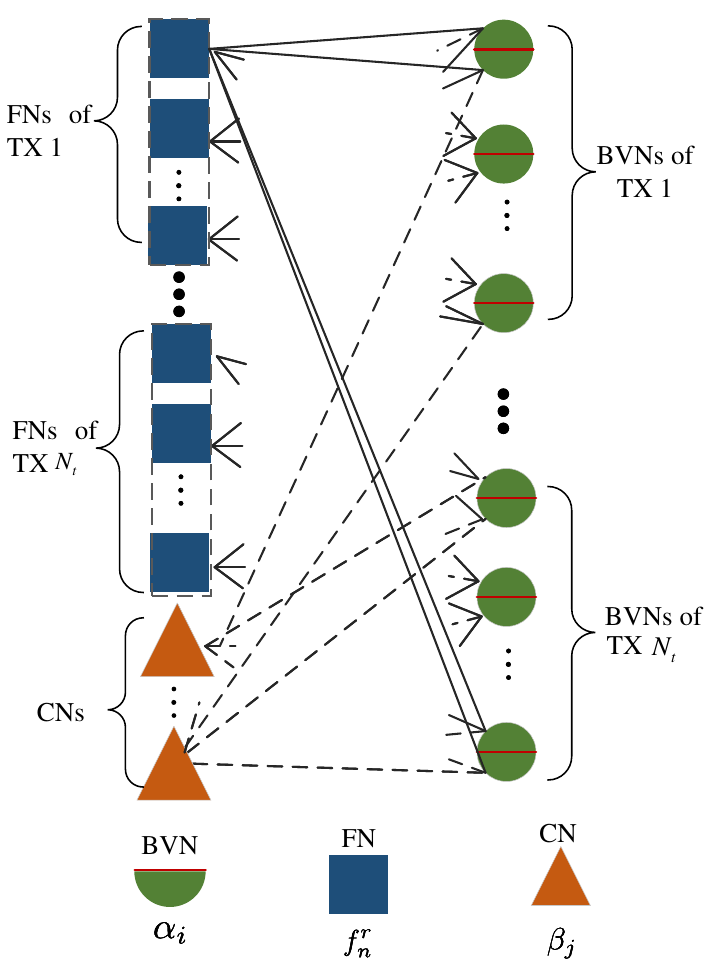}
     \caption{ {Factor representation of the proposed E-JSG.}}
     \label{Factor_ejsg}
         \vspace{-1em}
 \end{figure}
  \vspace{-0.4em}
\subsection{The Proposed E-JSG}

     \vspace{-0.4em}

\subsubsection{Linear encoding-inspired VI graph design} Let us consider the linear encoding for a modulation process in the following way \cite{SCMAE2E2}
\begin{equation}
      {x} =\mathbf {g} \mathbf {u},
\end{equation}
where $\mathbf {g} =[ {g}_{1}, {g}_{2},\ldots, {g}_{\log _{2}M}]$ denotes the constellation generator vector and $\mathbf {u} =[u_{1},u_{2},\ldots,u_{\log _{2}\,\,M}]^{T}\in \{1,-1\}^{\log _{2}\,\,M}$ stands for the  instantaneous input binary message vector. By  including all the $\mathbf {u} $'s  according to its corresponding integer values in ascending order, we obtain a $\log_2M \times M$ matrix $\mathbf U$. For example, when $M= 4$, we have 
\begin{equation}
\small
      \mathbf {U} =\left [{ \begin{matrix} -+-+\\ --++ \end{matrix} }\right], 
\end{equation}
  where ``$+$'' and ``$-$'' denote $+1$ and $-1$, respectively. Hence, the 
signal model of (\ref{SigMod1}) can be rewritten as 
\begin{equation}
\label{SigMod3}
\small
\begin{aligned}
     \mathbf y_{r} = & \sum_{t =1}^{N_t}  \mathbf H^{(r,t)} \mathbf G   \mathbf u_{t } +  \mathbf{w}_{r} 
      =  \sum_{t =1}^{N_t} \overline{ \mathbf H}^{(r,t)}     \mathbf u_{t } +  \mathbf{w}_{r},
\end{aligned}
\end{equation}
where $\mathbf u_{t } \in \{1,-1\}^{N\log _{2} M \times 1} $ denotes the instantaneous input binary message vector of an AFDM symbol at the $t$th TA,   $\mathbf G = \big [ \underbrace{\mathbf g ^{\mathcal T}, \cdots, \mathbf g ^{\mathcal T}}_{N}\big]^{\mathcal T} \in \mathbb C^{N \times  N\log_2 M}$ denotes the linear encoding matrix, and $\overline{ \mathbf H}^{(r,t)} = \mathbf H^{(r,t)} \mathbf G $ denotes the combined channel matrix. By utilizing (\ref{SigMod3}), the VI graph   designed  only contains bit VNs (BVNs) instead of the symbol VN.  Hereafter, the VNs in the E-JSG are referred to as the BVNs.    As a result, the LLR calculation for   bit to symbol is no longer required and the interleaver (deinterleaver)  can also be eliminated.   Fig.  \ref{Factor_ejsg} shows an example of  a folded view of  the proposed E-JSG,  where $M =4$ is considered.   Based on the JSG framework introduced in Section \ref{SecJSG}, both FNs and CNs are able to   update their messages concurrently. Subsequently, the BVNs of VI and LDPC subgraphs simultaneously  update  the  mean and  variance, and  LLR   to FNs and CNs, respectively.   Therefore, it is reasonable to place  FNs (rectangles) and CNs (triangles) on one side, while drawing BVNs (circles) on the other side.  Note that one circle contains $\log_2 M$ BVNs.

 \begin{algorithm}[t]
\caption{Message Propagation in the Proposed E-JSG}
\label{algorithm1}
\begin{algorithmic}[1]
\STATE  \textbf{Initialize :}   $L_{\beta_{j} \rightarrow \alpha_{i  }} =0$, $ L_{f_n^r \rightarrow \alpha_{i}} =0$, $L_{\alpha_{i} \rightarrow \beta_{j  }} =0$, $ L_{\alpha_{i} \rightarrow  f_n^r} =0$,  $ \mu _{n,r\rightarrow i}^{0}=0$,   $ \xi _{n,r\rightarrow i}^{0}=10$,   $ \mu _{i\rightarrow n,r}^{0} = 0$,  $ \xi _{i\rightarrow n,r}^{0}=10 $, and $\forall i, \forall j, \forall n,  \forall r $.
\FOR{$l=1:L_t$}
\STATE  \textbf{Step 1  : BVNs update }  
\STATE   Update   $L_{  \alpha_{i}  \rightarrow \beta_{j}} $  based on (\ref{LAB})  \\
\STATE   Update   $ L_{ \alpha_{i}}   $  based on (\ref{LAF})  \\
\STATE   Update the mean and variance at the BVN, i.e.,   $ \xi _{i\rightarrow n,r}^{l} $  and $ \mu _{i\rightarrow n,r}^{l}$,  according to  (\ref{m_v_ejsg})  \\
\STATE  \textbf{Step 2 :  FNs and CNs update } 
\STATE    Update $ L_{ \beta_{j}  \rightarrow \alpha_{i}}$ according to (\ref{CNupd1}) 
\STATE    Update $ \mu _{n,r\rightarrow i}^{l}$ and $ \xi _{n,r\rightarrow i}^{l}$  according to (\ref{Mw_va}) 
\STATE    Update $   L_{f_n^r \rightarrow \alpha_{i}} $    according to (\ref{F_V_EJSG}) 
\ENDFOR
\STATE  \textbf{Output :  $L_{ \alpha_{i}} $ } 
\end{algorithmic}
\end{algorithm}

In the following, we present  the message propagation in the proposed E-JSG.  

\subsubsection{Update of    BVNs to CNs and FNs}

 The updating of  $i$th BVN $\alpha_{i} $ to the $j$th CN $ \beta_{j}$ can be calculated as  
\begin{equation}
\small
\label{LAB}
    L_{  \alpha_{i}  \rightarrow \beta_{j}} =  \sum_{(r,n) \in V(i)}  L_{f_n^r \rightarrow \alpha_{i}}   + \sum_{j ^{\prime} \in \eta_i  \setminus j } L_{\beta_{j^{\prime}} \rightarrow \alpha_{i  }}, 
\end{equation}
where $V(i)$ is the set of FNs connected to the $i$th BVN. Similarly, to update the BVNs to FNs, the BVNs gather  the information from both sides, i.e.,   
\begin{equation}
\small
\label{LAF}
    L_{ \alpha_{i}  }=  \sum_{(r,n) \in V(i)   
 }  L_{f_{n}^{r} \rightarrow \alpha_{i}}   + \sum_{i  \in \phi_j    } L_{\alpha_{i  } \rightarrow \beta_{j}}.
\end{equation}
In this paper, we consider  $M=4 $ with $\mathbf g = [1, -1j]$.  Then, the calculation of  mean and variance of the $i$th BVN  can be simplified as 
\begin{equation}
\small
\begin{aligned}
       \mu _{i}^{l}& = 1- 2p( \alpha_i^{l} =  -1 \vert \mathbf y)  = \frac { 1- \exp \left (  L_{ \alpha_{i}  } \right)}{{1 + \exp \left (   L_{ \alpha_{i} } \right)}}, \\
       \xi _{i}^{l} & =  \sum _{   a_{i} \in \{-1,1\} } p( \alpha_i^{l}= a_i \mid \mathbf y){\left |{a_i-\alpha_i^{l}}\right |}^{2}   = \alpha_i^{l} (\alpha_i^{l}-1)^2.
\end{aligned}
\end{equation}
One has the the mean  and variance  passing from VNs to FNs respectively as follows
\begin{equation}
\small
\label{m_v_ejsg}
\begin{aligned}
        \xi _{i\rightarrow n,r}^{l}=&\left ({\frac {1}{\xi _{i}^{l} }-\frac {1}{\xi _{n,r\rightarrow i}^{l-1}}}\right)^{-1}, \\
        \mu _{i\rightarrow n,r}^{l}=& \xi _{i\rightarrow n,r}^{l} \left({\frac {\mu _{i}^{l} } { \xi _{i}^{l}}  -  \frac { \mu _{n,r\rightarrow i}^{l-1}} {\xi _{n,r\rightarrow i}^{l-1} 
 }}\right).
\end{aligned}
\end{equation}
\subsubsection{Update of CNs and FNs to BVNs}  The update for  CNs   are   the same as that  in  (\ref{CNupd1}) and (\ref{CNupd2}).   For the mean and variance update in the opposite direction, similar to   ({\ref{eq_a}})-({\ref{eq_c}}) and (\ref{qnn}), we have 
\begin{equation}
\small
\label{Mw_va}
     \mu _{n,r\rightarrow i}^{l} =\frac {y_{n}^{r}-\overline Z _{n,r\rightarrow i}^{l} }{\overline h_{n,r}^{i}}, \; \xi _{n,r\rightarrow i}^{l}=\frac {N_0+ \overline B_{n,r\rightarrow i}^{l}}{\big |\overline h_{n,r}^{i}\big |^{2}},
\end{equation}
where $\overline h_{n,r}^{i}$ is the effective channel coefficient from the $i$th BVN to the FN $f_n^r$ as specified in  (\ref{SigMod3}), and  
\begin{equation}
\small
\label{zb1}
\begin{aligned}
   \overline Z _{n,r\rightarrow i}^{l} =& \sum _{ \substack {i \prime\in \overline F(n,r), \\ i \prime \neq i }} \overline h^{i \prime}_{n,r} \mu _{i \prime  \rightarrow n,r}^{l-1},  \\
   \overline B_{n,r\rightarrow i}^{l}=& \sum _{ \substack {i \prime\in \overline F(n,r), \\ i \prime \neq i}} \vert \overline h^{i \prime}_{n,r} \vert ^2\xi _{i \prime  \rightarrow n,r}^{l-1}.
\end{aligned}
\end{equation}

Finally, the posterior probabilities of the transmitted symbol in LLR domain is given by
\begin{equation}
\small
\label{F_V_EJSG}
\begin{aligned}
   L_{f_n^r \rightarrow \alpha_{i}} & =  \frac{-\left |{1+  \mu _{n,r\rightarrow i}^{l} } \right | ^{2} +\left |{1-  \mu _{n,r\rightarrow i}^{l} } \right | ^{2}}  {\xi _{n,r\rightarrow i}^{l}}  = -\frac{4 \mu _{n,r\rightarrow i}^{l}}  {\xi _{n,r\rightarrow i}^{l}}.
\end{aligned}
\end{equation}

The overall message propagation of the proposed E-JSG is summarized in Algorithm (\ref{algorithm1}).

\vspace{-0.8em}
 \subsection{Sparse Channel  Based on eSNR  }
\label{Sparseejsg}

In the proposed   JSG and E-JSG,   the computational complexity depends crucially on the degree at each FN,  which is determined by    $P$  and $\kappa_{\nu}$. Specifically, the degree at each FN of E-JSG is upper bounded by $P\kappa_{\nu}\log_2(M)$.  { In general, $P$ and $\kappa_{\nu}$ tend  to be relatively large, particularly in rich scattering scenarios and when fractional Doppler shift is significant.}   To reduce the computational complexity of the    {JSG and E-JSG,} one possible approach is to reduce the number of edges  of the  {VI graph},  { i.e., select a small value of  $\kappa_{\nu}$ for each  $ {\overline{\mathbf{H}}_p}$, thereby  reducing unnecessary message propagation.}  This is reasonable as some of the   non-zero elements in ${\mathbf{H}}$ has very small amplitude, the message propagating over the corresponding edges may not significantly  improve the error rate performance.  { To this end, we propose a  sparse channel method based on eSNR to determine  the  nonzero elements  for  message passing, which is  equivalent to determine $k_{\nu}$,   and to  approximate    multiple graph edges with
insignificant channel coefficients as a single edge on the VI
graph. }

\begin{figure}
    \centering
\includegraphics[width=0.55\linewidth]{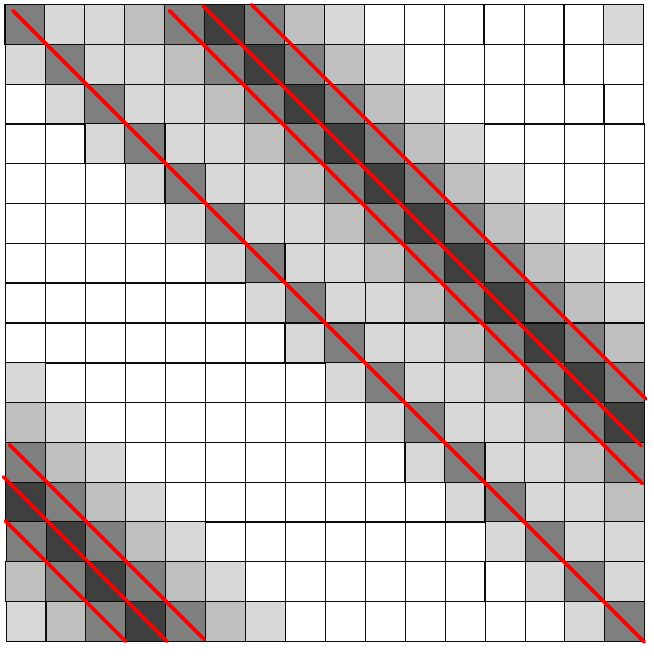}
    \caption{  {An example
of the selected entries,  highlighted  by the red lines,  for message propagation of a two-path channel. }}
    \label{channle_clp}
    \vspace{-1.5em}
\end{figure}

Denote the set of non-zero channel coefficients at the FN $f_n^r$ as $\mathcal D(n,r)$, and let $D= \vert \mathcal D(n,r)\vert$ return  the  cardinalities of  $\mathcal D(n,r)$. Then, we sort the $D$ non-zero channel coefficients in descending
order and choose the $D_{\text{L}}$ largest terms, denoted as $D_{\text{L}}(n,r)$,  for message propagation over the VI graph. We further denote  $D_{\text{S}}(n,r)$ as the   remaining terms. Then, the edges corresponding  to the $D_{\text{S}}(n,r) $ are    {approximated as a single edge.}   Consequently, the means and variance in (\ref{Mw_va}) and (\ref{zb1})  are computed respectively as follows
\begin{equation}
\small
\label{buld}
\begin{aligned}
   \overline Z _{n,r\rightarrow i}^{l} =& \sum _{ \substack {i \prime\in D_{\text{L}}(n,r), \\ i \prime \neq i }}   \overline h^{i \prime}_{n,r} \mu _{i \prime  \rightarrow n,r}^{l-1} + \mu _{\text{S} \rightarrow  n,r}^{l-1} \sum _{ \substack {i \prime\in D_{\text{S}}(n,r) }}  \overline h^{i \prime}_{n,r} ,  \\
   \overline B_{n,r\rightarrow i}^{l}=& \sum _{ \substack {i \prime\in D_{\text{L}}(n,r), \\ i \prime \neq i}} \vert \overline h^{i \prime}_{n,r} \vert ^2\xi _{i \prime  \rightarrow n,r}^{l-1} +\xi _{\text{S} \rightarrow  n,r }^{l-1} \sum _{ \substack {i \prime\in D_{\text{S}}(n,r) }} \vert \overline h^{i \prime}_{n,r} \vert ^2.
\end{aligned}
\end{equation}

  By   {approximating}  the insignificant  edges as a single edge, it allows more efficient message passing.    In what follows,  { we present the selection of $D_{\text{L}}(n,r)$ based on eSNR. }  
We first define the elements at the $n$th row and $m  = (n +s)_N$th column of ${ \mathbf H}^{(r,t)}$  as  the elements in its  $s$th cyclic-shift, where $0 \leq s \leq N-1$.  To maintain a regular structure of E-JSG, as per equation (\ref{SigMod1}),  all   elements at the $s$th  cyclic-shift  are either added to  {  $D_{\text{L}}(n,r)$    or  $D_{\text{S}}(n,r)$ for approximation.} 
   It should be noted that for each ${{\overline{\mathbf H}}_p}$, we have $\vert {{\overline{\mathbf H}}_p}[n,m] \vert = \vert {{\overline{\mathbf H}}_p}[n\prime,m \prime]\vert$, where $m  = (n + s)_N$,   $m  \prime= (n\prime +s)_N$, and $0 \leq s \leq N-1$.  However, this may not hold  for   ${ \mathbf H}^{(r,t)}$  as the phase of ${{\overline{\mathbf H}}_p}$ changes  along with the positions of $m  = (n +s)_N$,    leading to $\vert { \mathbf H}^{(r,t)}[n,m] \vert \neq \vert { \mathbf H}^{(r,t)}[n\prime, m \prime]\vert$ when the fractional Doppler and multipath  { exist}.    
  Hence, we define the average channel gain at the $s$th cyclic-shift as $G_s= \frac{1}{N} \sum_{n=1}^{N} \vert  { \mathbf H}^{(r,t)}[n,(n\prime +s)_N] \vert^2$.   The elements at the $s$th cyclic-shift  is added into $D_{\text{L}}(n,r)$ for message propagation   if and only if the eSNR of    the $s$th cyclic-shift is larger than a threshold, i.e,   
\begin{equation}
\small
\label{eSNR1}
  \text{eSNR}_s \equiv  10\log \frac{G_s}{N_0} \geq   \text{eSNR}_{\text{th}},  0\leq s \leq N-1,
\end{equation}
where $\text{eSNR}_{\text{th}}$ denotes the threshold, and   $ \text{eSNR}_s$ denotes the   {eSNR} at the $s$th cyclic-shift.  Fig. \ref{channle_clp} shows an example of the selected entries for message propagation of a two-path channel with $N=16$.    The elements at the $s = 0, 4,5 $ and $6$   cyclic-shifts are added to $D_{\text{L}}(n,r)$ for message propagation.

 \begin{table*}  
\small
\footnotesize
    \caption{Computational complexity}
    \centering
    \begin{tabular}{c|c|c|c|c|c}
    \hline
     \hline
       \textbf{ Algorithm}& Real-valued multiplications &Real-valued addition &  Exponentials &  $\text{Log}(\cdot)$ & \makecell{De-interleaving, \\interleaving ?} \\
        \hline
        \hline
                \makecell{ MMSE }& $6N^3+ 2N^2+2N$& -&    $2 N_V(\log_2 M +M)$& $2 N_V\log_2 M$ &Yes\\
                   \hline
                 \makecell{BP-JSG }&  $ 4 \sum_{n=1}^{N_{\text{F }}} {d_f^n}M^{d_f^n}$& \makecell{$2  \Big(\sum_{n=1}^{N_{\text{F }} } M^{d_f^n}+$\\ $ N_{\text{V}}(Md_v^{\text{ave}}  +   3\log_2 M -1)\Big)$} &   $2 N_V(\log_2 M +M)$     &$2 N_V\log_2 M$ & Yes\\
                  \hline
                 \makecell{EP-JSG }& $\makecell{4  (N_{\text{V}}(3M+8) +\\ N_{\text{F}}(3d_f^{\text{ave}}+3+2M))}$ & $\makecell{2 ( N_{\text{V}}(Md_{v}^{\text{ave}}+2M+3) +\\ N_{\text{F}}(2d_f^{\text{ave}}+2 +M) )} $ &         $2 N_V(\log_2 M +3M)$  &$2 N_V\log_2 M$  &Yes\\
                 \hline
    E-JSG& $N_V\log_2M(d_{v}^{\text{ave}}+7)$& $8N_V\log_2M + 6N_F\log_2Md_f^{\text{ave}} $ & $N_V\log_2M$ & - & No\\
  \hline
    \end{tabular}
    \label{Complexity}
        \vspace{-1em}
 \end{table*}


     \vspace{-0.8em}

\subsection{Complexity Analysis  }

In this subsection, we first analyze  the computational complexity of the proposed E-JSG.  We then discuss the decoding latency.

\textit{1) Computational complexity:}  The computation of  (\ref{FN_up}) requires   $\sum_{n=1}^{N_{\text{F}}}2{d_f^n}M^{d_f^n}$   complex-valued multiplications and $\sum_{n=1}^{N_{\text{F}}} M^{d_f^n}$ complex-valued additions  at each iteration, where $d_f^n$ denotes the number edges connected to the $n$th FN and $N_F $ denotes the number of FNs. In addition, the calculation of (\ref{VN_up}) requires $  N_{\text{V}}(Md_v^{\text{ave}} -1)$ complex-valued additions, where $d_{v}^{\text{ave}}=\frac{1}{N_V} \sum_{m=1}^{N_V}d_v^m $ denotes the average degrees of the VNs and $N_V $ denotes the number of VNs. Note that the output symbol LLRs are further demaped to bit-level LLRs based on (\ref{LLR1}), which requires $2 N_{\text{V}}\log_2 M$ real-valued additions,  $ N_{\text{V}}M$ exponentials,  and $  N_{\text{V}}\log_2 M$  $\text{log}(\cdot)$ calculations.    
The bit-level  LLRs to symbol level LLRs of (\ref{bit2sym}) requires  about $2\log_2 M $  exponentials,  $\log_2 M$ $\text{log}(\cdot)$ calculations and $2\log_2 M$  additions.

The messaging passing of VI graph with EPA principle  has linear computational complexity
with most of the key parameters, i.e., $M $, $ N_t$, $N_r$ and $N$. The update of mean and variance at from AVN to FN requires about $Md_{v}^{m}+2M+3$ additions, $2M$   exponentials, $3M+8$   multiplications at the $m$th AVN. Moreover, the update of mean and variance   from  the $n$th FN to AVN requires about $2d_f^{n}+2 +M$ additions, $3d_f+3+2M$ multiplications. Similarly, one can obtained the computational complexity for the   proposed E-JSG. The computational complexity of different receivers are summarized in Table \ref{Complexity}, where $d_{f}^{\text{ave}} = \frac{1}{N_F}\sum_{n=1}^{N_F} d_f^n    $. Since the complexities of the LDPC decoding are the same for different receivers, they are thus omitted in Table \ref{Complexity}.

\textit{2) Detection and  decoding latency:} 
  In practical systems, the detection and  decoding latency are mainly dominated by the iterative structures of the receiver.  We assume that the receiver is equipped with powerful  processing units, and   parallel computation techniques are also applied to reduce the detection and decoding latency.   The conventional MMSE with LDPC decoder, denoted as MMSE-LDPC, and the turbo receiver with iterative detection and decoding (Turbo-IDD)  are employed as the benchmarks for comparisons.  In the Turbo-IDD structure,  iterative messages are exchanged between the   detector  and channel decoder \cite{EnhancingLUo}.    Denote $T_{\text{VI}}^{\text{VN}}$ and  $T_{\text{VI}}^{\text{FN}}$ as the average run time for the  messaging propagation at the VNs and FNs of the VI graphs, respectively. Similar, let $T_{\text{LDPC}}^{\text{VN}}$ and  $T_{\text{LCPD}}^{\text{CN}}$ be the average run time for the  messaging propagation at the VNs and CNs of the LDPC graph, respectively.  Therefore, the detection and  decoding latency for the MMSE-LDPC receiver can be approximated as 
\begin{equation}
\small
\label{mmseldpc1}
 T_{\text{MMSE-LDPC}} =   T_{\text{MMSE}} +   N_{\text{LDPC}}(T_{\text{LDPC}}^{\text{VN}} +  T_{\text{LDPC}}^{\text{CN}})+ T_{\text{MMSE}}^{\text{res}},
\end{equation}
 where  $N_{\text{LDPC}}$ denotes the number of iterations of the LDPC decoding, and $T_{\text{SSD}}^{\text{res}}$ denotes the   residential run time, such as the symbol to LLR transformation and interleaving.  Similarly, the  detection and  decoding latency for the Turbo-IDD receiver can be expressed as
\begin{equation}
\small
 T_{\text{IDD}} =   N_{\text{out}}(N_{\text{VI}}(T_{\text{VI}}^{\text{VN}} +  T_{\text{VI}}^{\text{FN}}) +   N_{\text{LDPC}}(T_{\text{LDPC}}^{\text{VN}} +  T_{\text{LDPC}}^{\text{CN}})+ T_{\text{IDD}}^{\text{res}}),
\end{equation}
where $N_{\text{out}}$ denotes the outer iteration of the Turbo-IDD receiver, $N_{\text{VI}}$ and $N_{\text{LDPC}}$ denote the number of iterations of the VI detection and LDPC decoding, respectively, and $T_{\text{IDD}}^{\text{res}}$ denotes the residential run time. In addition, the  detection and  decoding latency for the proposed VI-JSG and E-JSP can be respectively expressed as 
 \begin{equation}
\small
 T_{\text{JSG}} =   N_{\text{JSG}}\left( \max\left \{ T_{\text{VI}}^{\text{VN}},  T_{\text{LDPC}}^{\text{VN}} \right\} + \max\left \{ T_{\text{VI}}^{\text{FN}},  T_{\text{LDPC}}^{\text{CN}} \right\} +      T_{\text{JSG}}^{\text{res}}\right),
\end{equation}
 and
 \begin{equation}
\small
\label{ejsg}
 T_{\text{E-JSG}} =   N_{\text{E-JSG}}\left( \max\left \{ T_{\text{VI}}^{\text{VN}},  T_{\text{LDPC}}^{\text{VN}} \right\} + \max\left \{ T_{\text{VI}}^{\text{FN}},  T_{\text{LDPC}}^{\text{CN}} \right\} \right).
\end{equation}
 where $N_{\text{JSG}}$ and $N_{\text{E-JSG}}$ denote  the number of iterations for the proposed JSG and E-JSG receivers, respectively.

\begin{figure}
    \centering
    \includegraphics[width=0.75\linewidth]{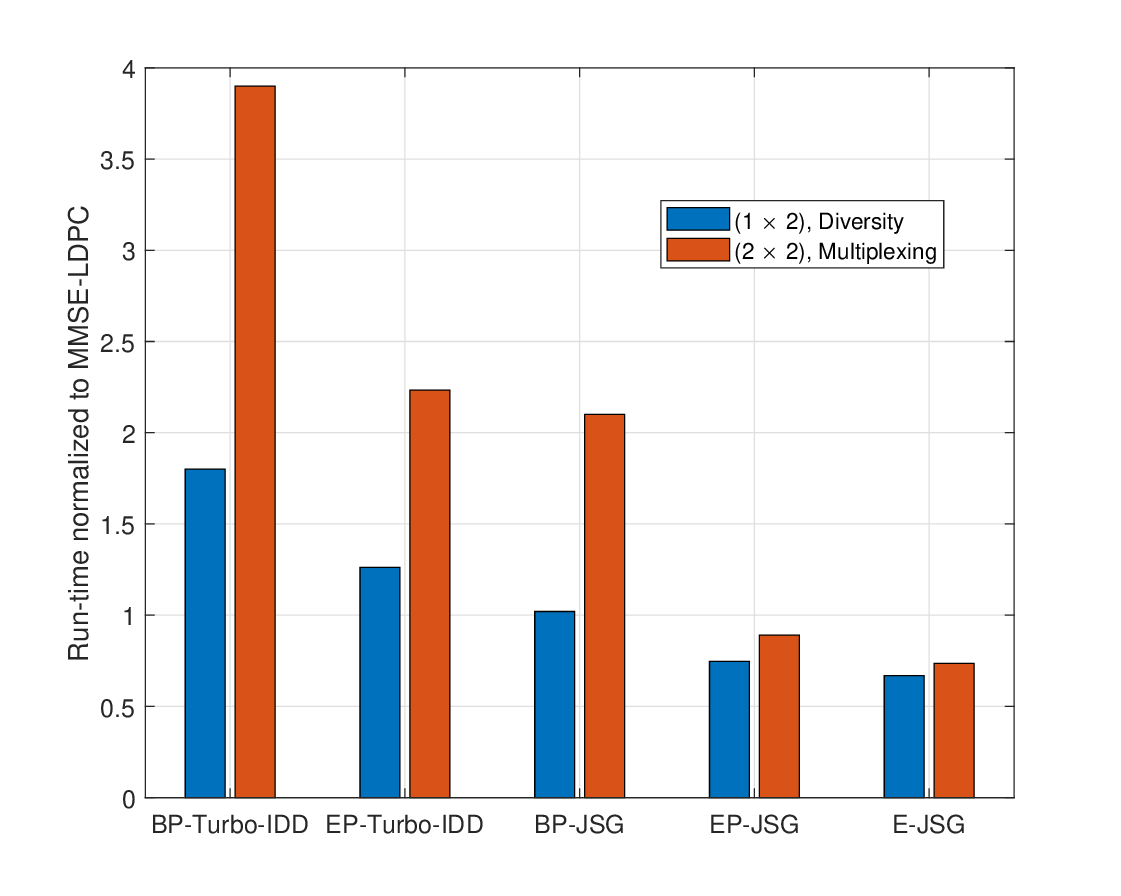}
    \caption{Normalized run-time complexity to MMSE-LDPC. }
    \label{runtime}
         \vspace{-1em}
\end{figure}
 

\section{Numerical Results}

\label{Sim}

\begin{figure*}[htbp]
	\centering
	\begin{subfigure}{0.32 \textwidth}
  \includegraphics[width=0.99 \linewidth]{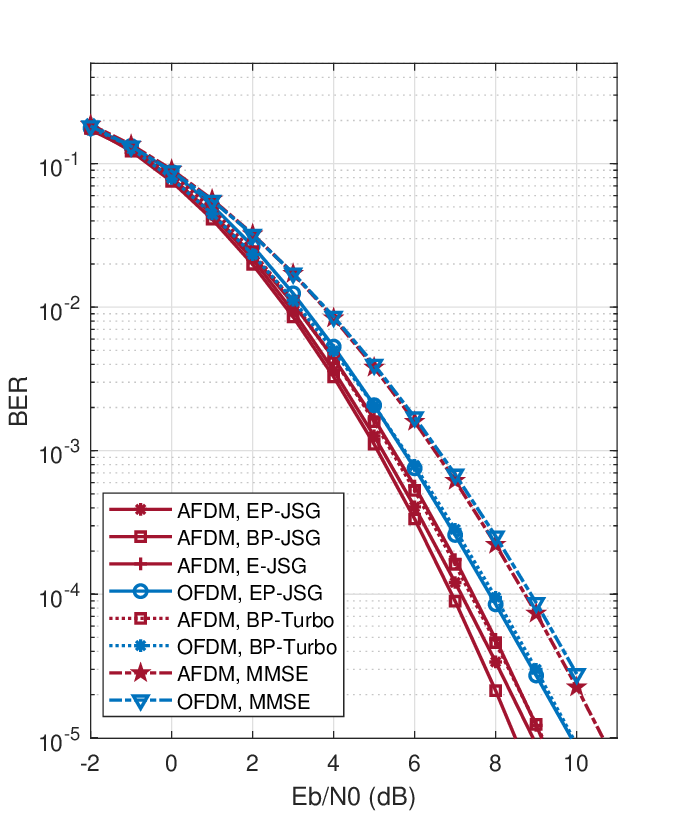}
		\caption{ $N_t = 1$, $N_r = 1$ and $P = 2$. }
				\vspace{-0.1em}
	\end{subfigure}
	\begin{subfigure}{0.32\textwidth}
  \includegraphics[width=0.99  \linewidth]{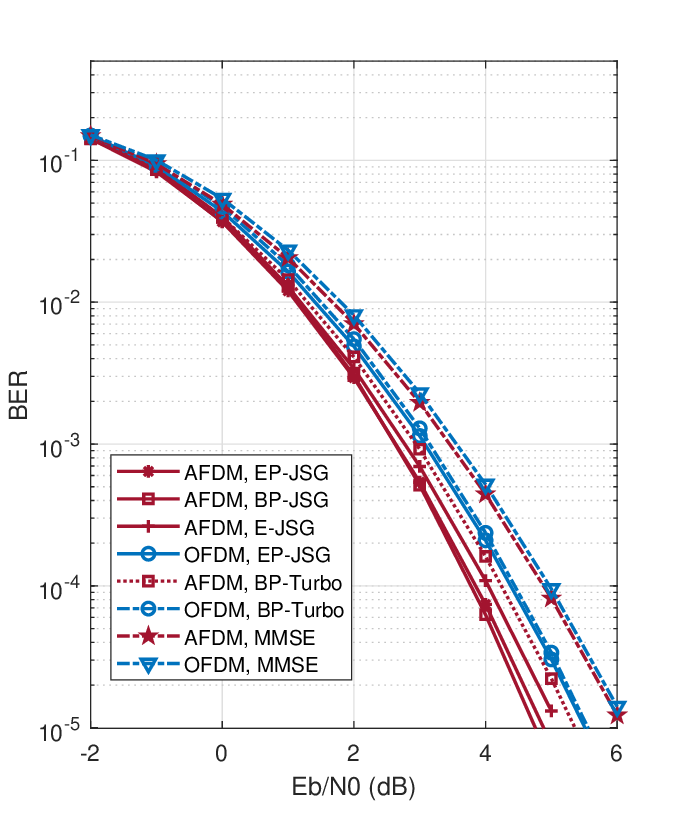}
		\caption{$N_t = 1$, $N_r = 2$ and $P = 2$.  }
		\vspace{-0.1em}
	\end{subfigure}
	\begin{subfigure}{0.32\textwidth}
  \includegraphics[width=0.99  \linewidth]{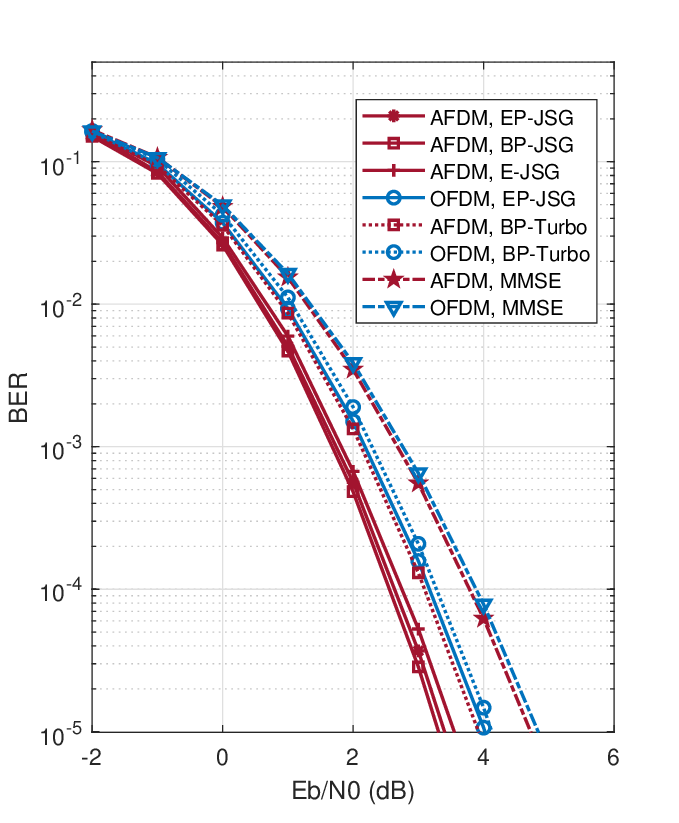}
		\caption{$N_t = 1$, $N_r = 2$ and $P = 4$.  }
		\vspace{-0.1em}
	\end{subfigure}	
	\caption{Coded BER performance comparisons of different receivers.}
	\label{SER_eq}
             \vspace{-1.5em}
\end{figure*}

This section presents  numerical simulation results to validate the performance of the proposed receivers in MIMO-AFDM systems.   {For system setting, we set   $N = 128 $, $c_1=c_2=\frac{1}{N}$, and  $N_{\text{CPP}} = 24$. Two maximum normalized Doppler values are considered, i.e., $\nu_{\max, p}=0.075 $ and $1.2$.
In addition,} we set $N_{\text{VI}} = 3$,  $N_{\text{LDPC}} =3$ and  $N_{\text{out}} =4$ for Turbo-IDD receivers.  The number of iterations for both the proposed JSG and the channel decoder in the MMSE-LDPC is set to $12$.  
The 5G NR LDPC code with a length of $512$ and rate of $0.5$ is considered. Specifically,  base graph $2$ is used, with LVN  and  CN degree distributions given by $\lambda(x) = 0.4x^2 + 0.4x^3 + 0.2x^5
$ and $\rho(x) = 0.5x^6 + 0.5x^7$, respectively,  as specified in 3GPP TS-$38.212$ \cite{3gpp5gnr}.

We first evaluate the run-time complexity as the approximation of the detection and decoding latency of various receivers. Specifically, the   run-time parameters of $T_{\text{MMSE}}^{\text{res}}$ in   (\ref{mmseldpc1})-(\ref{ejsg}) are estimated in the Matlab platform with an  i$7$ processor. Fig. \ref{runtime} shows the normalized run-time complexity of MMSE-LDPC of different receivers, i.e., $T_{\text{IDD}} / T_{\text{MMSE-LDPC}}$,  $T_{\text{JSG}} / T_{\text{MMSE-LDPC}}$ and $T_{\text{E-JSG}} / T_{\text{MMSE-LDPC}}$.  As can be seen from the figure, the BP-Turbo-IDD receiver exhibits the highest run-time complexity, whereas the proposed E-JSG receiver benefits from the lowest run-time complexity.

We now evaluate the performance of MIMO-AFDM with different receivers over multi-path fading channels. Specifically, we consider a two-path and a four-path channel without Doppler effects, where the delay taps are assumed to be randomly located within $[0, 6]$ and each path has a power of $1/P$.   Fig. \ref{SER_eq} shows the coded BER performance in SISO and $(1 \times 2)$-AFDM systems. The
main observations are summarized as follows: 1)  AFDM achieves a   BER performance similar to  that of OFDM systems when a conventional MMSE-LDPC receiver is applied. This is reasonable   as coded OFDM can also exploit multipath diversity, as pointed out in \cite{OCDMPerformance};

  2)  The proposed receivers achieve  about $2$ dB gain over the MMSE-LDPC receiver,  where the gain is more prominent as   $P$ and $N_r$ increase. This indicates that the proposed receivers can better exploit the diversity gain of MIMO-AFDM systems than the MMSE-LDPC receiver. In addition, the proposed receivers also achieve better coded BER performance than the turbo-IDD receivers; 3) It is interesting to observe that the BP-JSG achieves a similar BER performance to   EP-JSG. 
  In addition, the proposed E-JSG has a  slightly performance degradation compared to that of the EP-JSG, where the gap becomes negligible as   $P$ and $N_r$ increase.  { Similar observations have also been reported in sparse code  multiple access systems with turbo-like receivers \cite{yuan2018iterative}. This indicates that using both $\text{KL} ( p \vert q) $ and $ \text{KL} ( q \vert p)$ in the proposed JSG can well approximate the \textit{a posterior} probability of $\mathbf{x}$.   }

\begin{figure}[htbp]
	\centering
	\begin{subfigure}{0.45 \textwidth}
  \includegraphics[width=0.99 \linewidth]{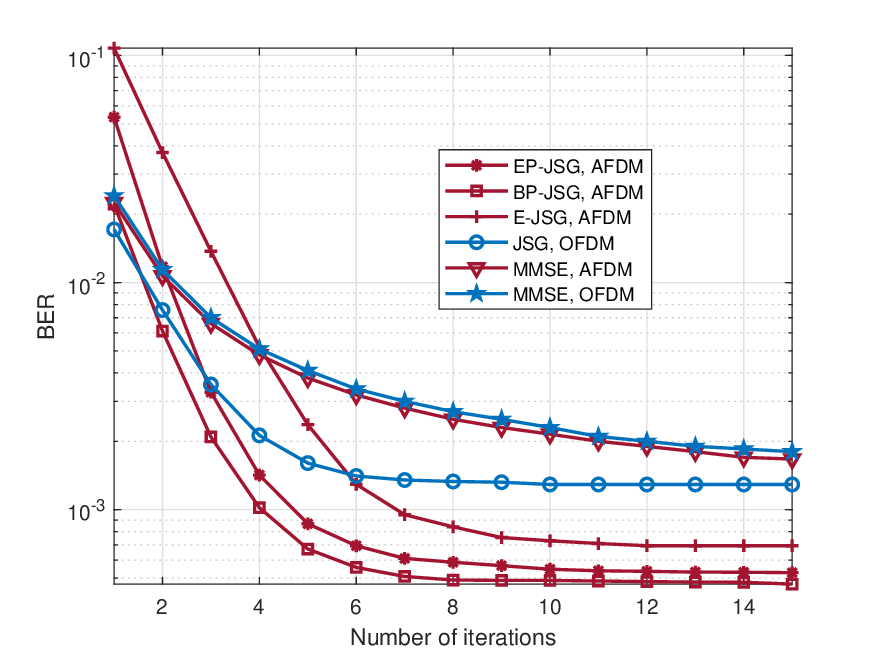}
		\caption{ {BER   v.s. Number of iterations. } }
				\vspace{-0.1em}
                \label{Converg1}
	\end{subfigure}
	\begin{subfigure}{0.45\textwidth}
  \includegraphics[width=0.99  \linewidth]{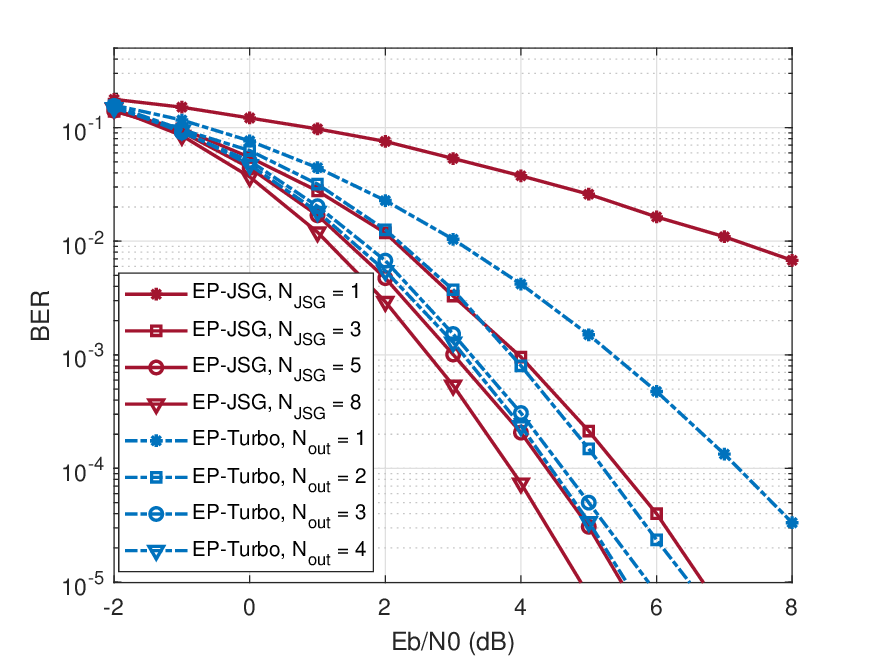}
		\caption{BER v.s. number of iterations ($P=2$). }
		\vspace{-0.4em}
        \label{Converg2}
	\end{subfigure}
	\caption{Convergence behavior comparisons of different receivers for $(1 \times 2)$ MIMO setting.}
             \vspace{-1.5em}
        \label{Converg}
\end{figure}

\begin{figure}
    \centering
\includegraphics[width=0.85\linewidth]{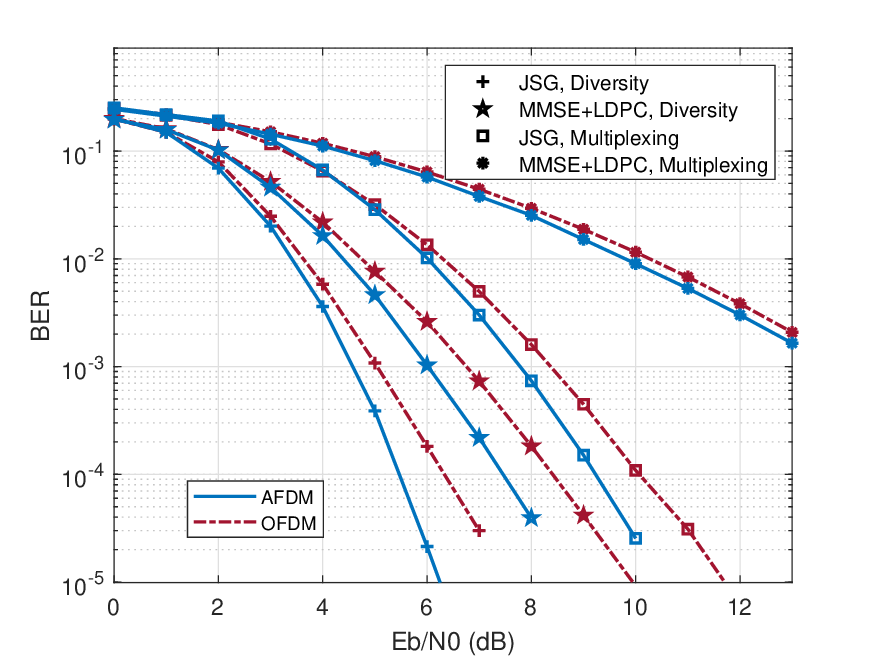}
    \caption{BER performance of ($2 \times 2$)-AFDM systems.}
    \label{22MIMO}
                 \vspace{-1em}
\end{figure}

 Fig.  \ref{Converg1} shows the BER performance with different number of iterations for various receivers in the MIMO-AFDM settings of $N_t = 1$, $N_r = 2$ and $P = 2$. As can be seen from the figure, the proposed BP-JSG and EP-JSG need about $7$ iterations to  converge, whereas  about $9$ and $11$ iterations are required for the proposed E-JSG and MMSE-LDPC receivers, respectively. It should be noted that the proposed E-JSG  allows a simultaneously detection and decoding at each iteration, and eliminates the need for interleavers, de-interleavers, symbol-to-bit, and bit-to-symbol LLR transformations.   
Fig.   \ref{Converg2} shows the coded BER performance at different iterations for the  JSG and turbo receivers. For simplicity, the EP-JSG and EP-turbo-IDD are considered. As can be seen from the figure, the EP-turbo-IDD receiver almost converges  after $3$ outer iterations. Namely, about $9$ EP and LDPC iterations   in total.  Moreover, consider the same number of EP and LDPC iterations, the proposed EP-JSG achieves better BER performance. For example, the proposed EP-JSG achieves about $2$ dB gain compared to that of the  EP-turbo-IDD receiver at the $\text{BER} = 10^{-4}$ for $3$ iterations, i.e,  $N_{\text{iter}}^{\text{out}}=1$.

\begin{figure} 
	\centering
	\begin{subfigure}{0.42 \textwidth}
  \includegraphics[width=0.99 \linewidth]{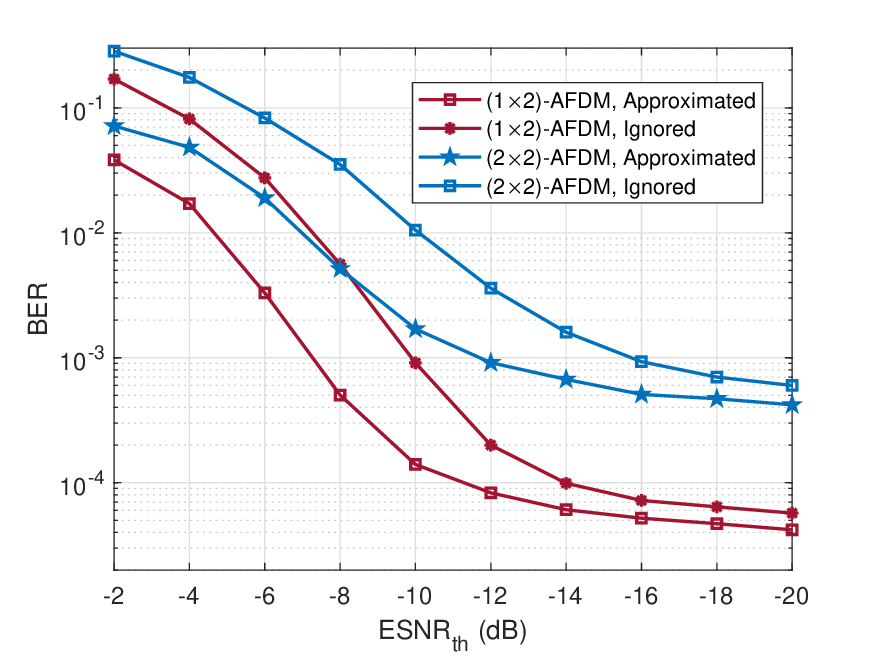}
		\caption{  {{BER performance v.s. $ \text{eSNR}_{\text{th}}$ ($P=4$). }} }
				\vspace{-0.1em}
	\end{subfigure}
	\begin{subfigure}{0.42\textwidth}
  \includegraphics[width=0.99  \linewidth]{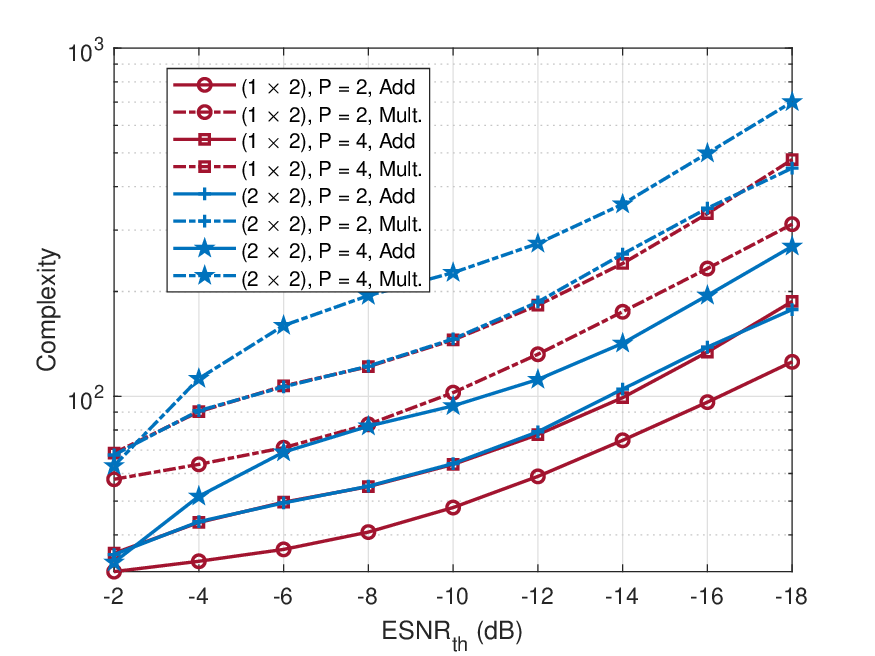}
		\caption{Complexity v.s. $\text{eSNR}_{\text{th}}$.  }
		\vspace{-0.4em}
	\end{subfigure}
	\caption{ {Coded BER performance  and complexity comparisons with different $ \text{eSNR}_{\text{th}}$ values. }}
	\label{complexity}
             \vspace{-1em}
\end{figure}

\begin{figure}
    \centering   \includegraphics[width=0.85\linewidth]{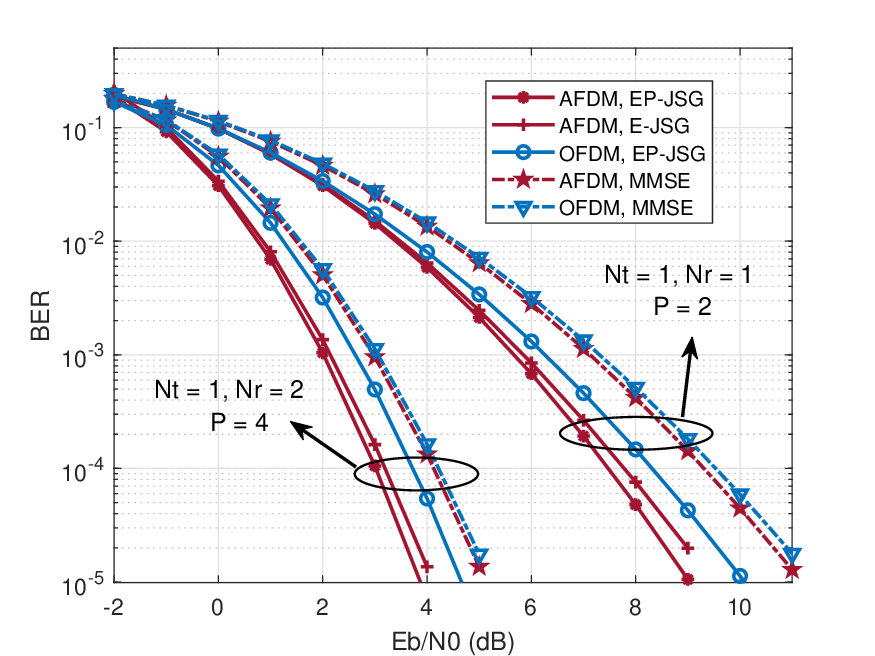}
    \caption{BER performance with Doppler effects.}
    \label{DopplerFig}
             \vspace{-1em}
\end{figure}
Fig. \ref{22MIMO} shows the coded BER performance in an  ($2 \times 2$)-AFDM system. Specifically, both   transmission scenarios $1$ and $2$ are considered. For multiplexing transmission of Scenario $1$, the MMSE-LDPC deteriorates significantly. Notably,  the proposed E-JSG for the diversity and multiplexing transmissions   achieve  about $2$ dB and $4$ dB gains, respectively, compared  to   the MMSE-LDPC receivers for $P=2$. In addition, compared to the ($1 \times 2$) MIMO setting in Fig. \ref{SER_eq}, the performance of ($2 \times 2$)-AFDM with diversity transmission shows no improvement. This occurs because the transmit antennas may share the same delay taps, leading to a sacrifice in diversity.

Next, we present the coded BER performance and computational complexity of the proposed sparse channel message propagation method based on eSNR,  which is shown in Fig. \ref{complexity}.   Fig.   \ref{complexity}(a) illustrates the coded BER performance with different eSNR thresholds. The curves labeled  {``Approximated" } represent the proposed scheme, whereas ``Ignored" indicates that the insignificant edges in $D_{\text{L}}(n,r)$  are ignored. Clearly, when these insignificant edges are ignored, the performance deteriorates significantly, particularly for small values of $\text{eSNR}_{\text{th}}$.
It is also observed that the performance of the proposed E-JSG deteriorates rapidly for $ \text{eSNR}_{\text{th}} > -12$ dB.     
Fig.   \ref{complexity}(b) shows the average number of addition  and multiply operations of each node at each iteration for different MIMO-AFDM settings.  Interestingly,  the computational complexity decreases almost linearly as  $ \text{eSNR}_{\text{th}}$ increases.  This  indicates that  as  $ \text{eSNR}_{\text{th}}$ increases,  the average  $d_f$ and $d_v$ at each FN and VN also decrease linearly.    { Consider the trade-off   between BER performance and computational complexity, we set $  \text{eSNR}_{\text{th}} = -12$ dB for both JSG and E-JSG.}

  \begin{figure}
    \centering
\includegraphics[width=0.86\linewidth]{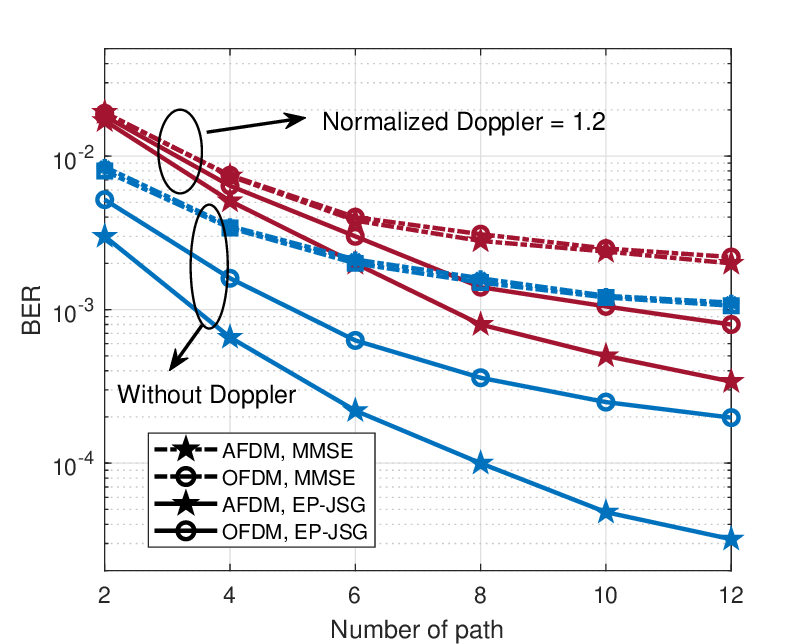}
    \caption{  {Coded BER performance v.s. number of path in ($1 \times 2$)-AFDM and -OFDM systems at the $\text{E}_\text{b}\/\text{N}_0 = 2$ dB.}}
    \label{BER_P}
             \vspace{-1.5em}
\end{figure}

 {Fig. \ref{DopplerFig} shows the coded BER performance of MIMO-AFDM in a fast time-varying channel  with the maximum normalized Doppler $\nu_{\max} = 0.075$ and  $  \text{eSNR}_{\text{th}} = -12$ dB.   The proposed E-JSG  and   MMSE-LDPC receivers are considered as benchmarks  for comparison. The normalized Doppler shift  at the $p$th   path    is  given by $\nu_{p} = \nu_{\max}\cos(\psi_{p})$ where $\psi_{p}$ has a uniform distribution of $\psi_{p,i} \sim \mathcal U [-\pi, \pi]$.   
The proposed sparse channel scheme is applied to both JSG and E-JSG.}  
As observed from the figure, the proposed E-JSG receiver exhibits approximately a $2$ dB improvement over the MMSE-LDPC receiver at the BER of $10^{-5}$ for SISO-AFDM with $P=2$. For the $(2 \times 2)$-AFDM with $P=4$, the gain is approximately $1$ dB. Additionally, there is a performance degradation of about $0.7$ dB for the proposed receiver compared to the MIMO-AFDM with zero Doppler.

 { Fig. \ref{BER_P} further demonstrates the coded BER performance in OFDM and AFDM systems  as   the number of paths increases by considering a high-mobility channel with severe  Doppler effect, i.e., $\nu_{\max} = 1.2$.  Specifically, we consider the MMSE and EP-JSG receivers, and a $(1 \times 2)$ MIMO setting. We assume each path has the same power, and $N_{\text{cpp}}$ is larger than the maximum integer delay. Both AFDM and OFDM with MMSE and EP-JSG perform better as $P$ increases, indicating  that AFDM and OFDM with MMSE and EP-JSG receivers are able to collect the diversity gain even in the presence of severe Doppler effect ($\nu_{\max} = 1.2$).  Moreover, the gap between AFDM and OFDM with the proposed EP-JSG becomes smaller for $\nu_{\max} = 1.2$ compared to the case with no Doppler.    This is reasonable, as more channel elements in the AFDM system   are approximated by the proposed sparse channel scheme for $\nu_{\max} = 1.2$   compared to the case of $\nu_{\max} = 0$. However, AFDM with the proposed JSG receiver  still achieves better BER performance than OFDM systems or MMSE receivers. }

\section{Conclusion}
\label{conclu}
In this paper, we have proposed JSG receivers for MIMO-AFDM systems.  
Specifically, the BP and EP as detection techniques have been introduced based on a unified VI perspective. In addition,   by representing the VI graph and the LDPC codes as bipartite graphs, we have constructed a JSG for MIMO-AFDM, which enables  simultaneous detection and decoding.  We delve into the detailed messaging propagation over the proposed JSG. In addition, we have further  proposed an  E-JSG  receiver based on a linear constellation encoding model, which can eliminate the need for interleavers, de-interleavers, and LLR transformations, thereby enhancing detection and decoding efficiency. Additionally, we have introduced a sparse channel method to reduce detection complexity. Simulation results demonstrated the superiority of our approach, in terms of  computational complexity,   latency, and   error rate performance over conventional receivers. We have showed that the proposed JSG receivers efficiently exploit multipath and Doppler diversity, thereby outperforming conventional MMSE and turbo-IDD receivers, and demonstrating the superiority of AFDM over OFDM.


%





\ifCLASSOPTIONcaptionsoff
  \newpage
\fi



%

\bibliography{ref} 
\bibliographystyle{IEEEtran}

 %








\end{document}